\documentclass[aps,pra,superscriptaddress]{revtex4}
\usepackage[intlimits]{amsmath}
\usepackage{amsfonts}
\usepackage{graphics}
\usepackage{bbm}
\usepackage{subfigure}
\usepackage[usenames]{color}
\usepackage{epstopdf,epsfig}
\usepackage{mathtools}
\usepackage{ulem}
\usepackage{soul}

\advance\voffset by -1cm

\DeclarePairedDelimiterX\braket[2]{\langle}{\rangle}{#1 \delimsize\vert #2}

\usepackage{indentfirst}

\usepackage{graphicx}
\usepackage{amsmath}
\usepackage{amsfonts}

\newcommand\be            {\begin{equation}}
\newcommand\bea           {\begin{equation}\begin{array}l\displaystyle}
\newcommand\ee            {\end{equation}}
\newcommand\bes           {\begin{subequations}}
\newcommand\esu           {\end{subequations}}

\makeatletter
\newcommand{\vast}{\bBigg@{3.3}}
\newcommand{\Vast}{\bBigg@{5}}

\makeatother

\newcommand\p            {\partial}

\def\3pt#1#2#3{{\langle{#1}\vert{#2}\vert{#3}\rangle}}
\def\barray{\begin{eqnarray}}
\def\earray{\end{eqnarray}}

\usepackage{tikz}
\usetikzlibrary{calc}

\begin{document}

\title{
  Dynamics of one--dimensional quantum many-body systems \\
  in time--periodic linear potentials  
}

\author{A. Colcelli}
\affiliation{SISSA and INFN, Sezione di Trieste, Via Bonomea 265, I-34136 
Trieste, Italy}

\author{G. Mussardo}
\affiliation{SISSA and INFN, Sezione di Trieste, Via Bonomea 265, I-34136 
Trieste, Italy}

\author{G. Sierra}
\affiliation{Instituto de F\'isica Te\'orica, UAM/CSIC, Universidad 
Aut\'onoma de Madrid, Madrid, Spain}

\author{A. Trombettoni}
\affiliation{Department of Physics, University of Trieste, Strada
  Costiera 11, I-34151 Trieste, Italy}
\affiliation{CNR-IOM DEMOCRITOS Simulation Center, Via Bonomea 265, I-34136 
Trieste, Italy} 
\affiliation{SISSA and INFN, Sezione di Trieste, Via Bonomea 265, I-34136
Trieste, Italy}

\begin{abstract}
  We study a system of one--dimensional interacting quantum particles subjected  to a time--periodic potential linear in space. After discussing the cases of
  driven one- and two-particles systems, we derive the analogous  results for the many-particles case in presence of a general interaction
  two-body potential and the corresponding Floquet Hamiltonian.  When the undriven model is integrable, the Floquet Hamiltonian is shown to be integrable too. 
  We determine the micro-motion operator   and the expression for a generic time evolved state of the system.
  We discuss various aspects of the dynamics of the system both at stroboscopic and intermediate times, in particular the motion of the
  center of mass of a  generic wavepacket and its spreading over time. We also discuss the case of accelerated
  motion of the center of mass, obtained when the integral of the coefficient strength of the linear potential on a time period
  is non-vanishing, and we show that the Floquet Hamiltonian  gets in this case an additional static linear potential. We also discuss the
  application of the obtained results  to the Lieb--Liniger model. 
\end{abstract}

\maketitle

\section{Introduction}
Time--periodic driven quantum systems have become recently the subject of
an intense research activity.
These out of equilibrium systems give rise to interesting novel physical
properties as, for instance, dynamic localization effects \cite{Dunlap1986},
suppression of tunneling
subjected to a strongly driven optical lattice
\cite{Creffield2003,Eckardt2005,Lignier2007,Kierig2008,Eckardt2009,Sierra2015,He2019} (see \cite{Eckardt2017} for more references),
topological Floquet phases \cite{Kitagawa2010,Lindner2011}, time crystals 
\cite{Wilczek2013,Watanabe2015,Choi2017,Zhang2017,Russomanno2017,Yao2017,SachaZak2018,Sacha2018}, 
dynamics in driven systems \cite{Russomanno2012,Goldman2014,Holthaus2016} and 
Floquet prethermalization \cite{Weidinger17,Herrmann18}.
All these concepts and phenomena can be collected together under the heading of 
``Floquet engineering'' \cite{Eckardt2017,Oka18}, a very active
field both from experimental and theoretical points of view. 

The name itself came from a famous 
paper by Floquet \cite{Floquet883}, who was interested
in the study of differential equations with coefficients given by
time--periodic functions. The formalism he developed turns out to
be very helpful in dealing with the Schr\"odinger equation of a
Quantum Mechanical  system with a time--periodic Hamiltonian
\cite{Shirley65,Grifoni98}. Preparing the system in
an initial state $\chi (t=0)$ and letting the periodic driving act on it,
the Floquet Hamiltonian $\hat{H}_F$ is the operator that formally gives the
state of the system at multiples of the period $T$:
\be
\chi(t=nT)=e^{-i\frac{nT}{\hbar}\hat{H}_F}\,\chi(t=0)\,.
\label{Fl_int}
\ee
In other words, the Floquet Hamiltonian $\hat{H}_F$
determines the stroboscopic evolution of the system.
It depends on the parameters of the original undriven Hamiltonian, $\hat{H}_0$, 
and on the time--dependent perturbation.
$\hat{H}_F$ is a hermitian operator whose eigenvalues are
the so called quasi-energies $\mathcal{E}_F$.
On the other hand, the evolution of the state $\chi(t=0)$
at generic times $t \in (0,T)$
is determined by the \textit{micro-motion operator} $\hat{U}_F(t,0)$,
defined by the following decomposition of the time evolution operator
of the system $\hat{U}(t,0)$:
\be
\hat{U}(t,0)=\hat{U}_F(t,0) \,e^{-i\frac{t}{\hbar}\hat{H}_F}\,.
\label{micro}
\ee
Applying the micro-motion operator $\hat{U}_F(t,0)$ on the
eigenstates of the Floquet Hamiltonian and multiplying by a complex
exponential containing the quasi-energies, one obtains the
\textit{Floquet states} $\left| \psi_F(t)\right\rangle$. They form a
complete and orthonormal set of functions and therefore any solution of the original time--dependent Schr\"odinger equation can be written as a superposition
in terms of them 
\be
\nonumber
\chi(t) = \int dk \, A(k) \, \left|\psi_F(t)\right\rangle\,,
\ee
where $k$ is a momentum variable, related to the energy of the system ($k\propto \sqrt{E}$), and the $A(k)$'s are time--independent coefficients. Therefore
finding $\hat{H}_F$ and $\hat{U}_F(t,0)$ gives access 
to the full quantum dynamics of the system.

Finding the Floquet Hamiltonian and the micro-motion operator for an interacting many-body system in the  presence of a time--dependent driving is in general
a challenging and highly interesting task, relevant for a variety of applications
in the field of Floquet engineering. Tuning the form and the
parameters of the undriven system and of the periodic perturbation, one aims at
controlling the (desired) effective Hamiltonian
of the quantum dynamics of the system itself.

In general, even if the undriven model is integrable, when we subject it to a time--periodic potential, we end up in a non-integrable Floquet
Hamiltonian. In a recent paper \cite{ourPRL2019} we addressed the question whether it would be possible
to have an integrable Floquet Hamiltonian by perturbing an integrable $1D$
bosonic model
with a time--periodic perturbation, finding  a positive  answer.
Namely, we considered
the integrable Hamiltonian that describes  a one--dimensional gas of bosons
with contact interactions, \textit{i.e.} 
the Lieb--Liniger Hamiltonian \cite{LiebLiniger1963}, in the presence of
a linear in space, time--periodic one-body potential of the form
\be
V(x,t)=f(t) \, x  \ , 
\label{pot_int}
\end{equation}
with a driving function $f(t)$ with period $T$: $f(t)=f(t+T)$. 
It was shown  in \cite{ourPRL2019}
that under the condition 
\begin{equation}
  \int_0^T f(\tau)\,d\tau=0\, \ , 
  \label{cond}
 \end{equation} 
the resulting Floquet Hamiltonian is integrable and 
has a Lieb--Liniger form, with a shift on the momenta of the particles.

Despite the fact that other exactly solvable time--dependent Hamiltonians
can be constructed using
different approaches \cite{Yuzbashyan18,Sinitsyn18},
the problem of finding an integrable
Floquet Hamiltonian from an undriven interacting one 
is in general a difficult task. The goal of the present paper is two-fold: (a) first, 
we provide a derivation valid for general
one--dimensional many-particles systems, extending the results 
of \cite{ourPRL2019} to an arbitrary two-body interaction potential
$V_{2b}(x_j-x_i)$ and giving explicit results for the micro-motion operator
$\hat{U}_F(t,0)$; (b) secondly, we present  a detailed
  discussion of the case in which the condition
  (\ref{cond}) does not hold, emphasizing its role for the time--dependence
  of the energy of the system. 
  
  We will show that if the undriven Hamiltonian is integrable
  and perturbed with a linear time--periodic potential,
  then also the Floquet Hamiltonian is integrable if the driving function
  has a vanishing integral over a period of oscillation, as it occurs
  for the Lieb--Liniger case.
  If, on the contrary, the condition (\ref{cond})
  does not hold, we will see that the Floquet Hamiltonian can be still recast in a time--independent expression 
   but with the addition of a linear potential. 
  Expressions for the value of the energy during the stroboscopic dynamics
  are found and the micro-motion operator explicitly written down.
  The method we use is based on first applying a gauge transformation
  on the wavefunction to wash out the linear term 
  and then solving the time--dependent
  Schr\"odinger equation with Hamiltonian $\hat{H}_F$.
  It is worth to underline that, in general, one of the difficulties
  in identifying integrable Floquet Hamiltonians is that the
  integrability of these Hamiltonians is not at all guaranteed by the
  integrability of the original time--independent undriven model
  (see, for instance, \cite{Komnik16} where starting from the original
  BCS model the corresponding BCS gap equation in the  presence of a
  periodic driving is derived and solved numerically).
  For the class of one--dimensional
  interacting many-particles systems considered here,
  we show instead that it is not the case, as far as the periodic driving
  is a linear function on the position variables.

  In the following we present a 
  detailed analysis of all these aspects of the problem and, in particular,
  we show how to extract the time evolution of generic wavefunctions at
  all times by first computing the micro-motion operators and then the
  Floquet states, with which we can expand the wavefunction. After discussing
  a general two-body interaction term, we focus on the
  paradigmatic and experimentally relevant case where the particles
  interact with contact interactions, \textit{i.e.} the Lieb--Liniger model. This model
  constitutes an ideal playground for integrable models
  in one--dimensional continuous space. It is indeed exactly
  solvable using Bethe ansatz techniques
  \cite{LiebLiniger1963,Yang1969,Korepin1993,Mussardo10}, related to the non-relativistic limit of the Sinh-Gordon model \cite{KTM} 
  and routinely used to describe (quasi-) one--dimensional bosonic gases
  realized in ultracold atoms experiments
  (see the reviews \cite{Yurovsky2008,Bouchoule2009,Cazalilla2011}).

  The paper is organised as follows. In order to set the notations and present initially 
  the general results in the simpler form, in Section II
  we discuss  the dynamics of the one-particle case,
  \textit{i.e.}
  the Schr\"odinger equation for a particle of mass $m$ in a linear
  time--periodic potential in one--dimension. In Section III we
  consider the interacting two-particles case, where both particles,
  in addition to their relative potential,
  are also subjected to a periodic driving potential proportional
  to their position.
  In Section IV, 
  we address the many-body interacting case. Our conclusions
  are finally gathered in Section V.  

  \section{One-body problem}
  \subsection{Generic driving function}
  \label{IIA}
Let us consider the one--dimensional Schr\"odinger equation
for a particle of mass $m$ in a linear potential with a time varying strength:  
\be
\label{onebody_schro}
i \hbar \frac{\p \chi}{\p t}=-\frac{\hbar^2}{2m} \frac{\p^2 \chi}{\p x^2} + x\,f(t)\,\chi(x,t)\,. 
\ee
In what follows,  $f(t)$ is a generic driving function  
that will be taken to be periodic at the end of this Section. 
In the literature, Eq. (\ref{onebody_schro}) has been studied and
solved in different ways \cite{Berry1978,Rau1996,Guedes2001,Feng2001}.
Here we solve it  with  a method that  will be particularly useful
to study  the Floquet dynamics. 

The key point of 
the solution of Eq. (\ref{onebody_schro}) is to perform a
gauge transformation on the wavefunction 
\be
\label{gauge_trasf_1b} 
\chi (x,t) = e^{i \theta(x,t)} \,\eta(y(t), t)\,,
\ee
where $y(t)=x-\xi(t)$, while $\xi(t)$ and $\theta(x,t)$ are 
two functions that are determined below.
Substituting Eq. (\ref{gauge_trasf_1b}) into  (\ref{onebody_schro}), 
and  imposing 
\be
\label{xi_condition}
\frac{d\xi}{dt}= \frac{\hbar}{m} \frac{\p \theta}{\p x}\,,
\ee
and 
\be
\label{theta_condition}
-\hbar \frac{\p \theta}{\p t} = \frac{\hbar^2}{2m} \left(\frac{\p \theta}{\p x}\right)^2 + xf(t)\,,
\ee
we find that  $\eta(y,t)$ satisfies the Schr\"odinger equation with no external
potential in the spatial variable $y$: 
\be
\label{schroeq_eta_1b}
i \hbar \frac{\p \eta}{\p t}= -\frac{\hbar^2}{2 m} \frac{\p^2 \eta}{\p y^2}\,. 
\ee
Hence, once $\theta(x,t)$ is known, $\eta(y,t)$
will be  readily determined from the free dynamics. 
To find the gauge phase $\theta(x,t)$ we make the ansatz
\be
\label{ansatz_theta}
\theta(x,t) = \frac{m}{\hbar} \frac{d\xi}{dt} \,x+ \Gamma(t)\,,
\ee
that leads to the conditions
\be
\label{vANDdelta_onebody}
m\frac{d^2 \xi}{dt^2} = -f(t)\,, \,\,\,\,\,\,\,\,\,\, \hbar \frac{d \Gamma}{dt}=-\frac{m}{2} \left(\frac{d\xi}{dt} \right)^2 \, , 
\ee
which  give the   translational parameter $\xi(t)$ and the function
$\Gamma(t)$ in terms of $f(t)$. Notice that the equation for  $\xi(t)$  is the Newton's second law
equation of motion, where $d^2 \xi / dt^2$ represents the acceleration of
the center of mass of the system, and $-f(t)$ the driving force.

Solving the equations  (\ref{vANDdelta_onebody}),  with the
initial conditions $\xi(0)\,= d\xi(0)/dt= 0$ and $\Gamma(0)=0$, we get 
\be
\label{theta_f(t)}
\theta(x,t)=-\frac{x}{\hbar} \int_0^t f(\tau)\,d\tau -\frac{1}{2\,m\,\hbar} \int_0^t d\tau \left[\int_0^\tau f(\tau')\,d\tau' \right]^2\,,
\ee
which, together with Eq. (\ref{gauge_trasf_1b}) and
Eq. (\ref{schroeq_eta_1b}), completely solves Eq. (\ref{onebody_schro}).

Since: $\theta(x,0)=0$
and $y(0)=x$, we have from Eq. (\ref{gauge_trasf_1b}) that
\be
\label{initial_conditions_wf}
\chi(x,0)=\eta(x,0)\,, 
\ee
for which the  solution of the Schr\"odinger equation
(\ref{onebody_schro}) reads 
\be
\label{complete_sol_1b}
\chi(x,t)\,=\,e^{i\theta(x,t)} e^{-i\frac{t}{\hbar}\frac{\hat{p}^2}{2m}}\,\eta(y,0)\,=\,e^{i\theta(x,t)} e^{-i\frac{t}{\hbar}\frac{\hat{p}^2}{2m}} e^{-i \frac{\xi(t)}{\hbar} \hat{p}}\,\chi(x,0)\,,
\ee
where we have used the definition of the translation operator and the
free time evolution operator. Notice that no boundary conditions in
the wavefunction have been considered in the above calculations.

In terms of the solution (\ref{complete_sol_1b}), one can easily compute
the expectation values of various physical quantities, such as momentum,
position as well as their variances. Assuming as initial values
$\left\langle \hat{x} \right\rangle (t=0) = x_0$ and
$\left\langle \hat{p} \right \rangle (t=0) = p_0$, and using the canonical commutation
relations among different powers of position and momentum operators, we have
\be
\label{x_mean_1b}
\left\langle \hat{x} \right \rangle(t) \equiv \left\langle \chi(x,t)\left| \hat{x} \right| \chi(x,t)\right\rangle = x_0 + \frac{t}{m} p_0 +\xi(t)\,.
\ee 
This means that the mean position of a generic wavepacket, 
under the action of a linear time--dependent potential, is governed by
the parameter $\xi(t)$ which is readily determined by Eq. (\ref{vANDdelta_onebody}).
Moreover, concerning the expectation value of the momentum we have 
 \be
\label{p_mean_1b}
\left\langle \hat{p} \right \rangle(t) \equiv \left\langle \chi(x,t)\left| \hat{p} \right| \chi(x,t)\right\rangle \,=\, p_0 -\int_0^t f(\tau)\,d\tau\,,
\ee 
meaning that the value of the momentum is shifted away 
 from its initial value by a term  that depends  on the driving function $f(t)$. As expected,  the
motion of the center of the wavepacket in Eq. (\ref{x_mean_1b}) is
the same of a classical particle moving  in one dimension
under the action of a time--dependent gravitational force.
Concerning  the variance of the position, we have 
\be
\label{variance_x_1b}
\Delta x(t) \equiv \sqrt{\left\langle \hat{x}^2\right\rangle(t)-\left\langle \hat{x}\right\rangle^2(t)} = \Delta x_{\rm undriven}(t)\,,
\ee
where the subscript "undriven" stands for the undriven 
evolution of the variance,
which is calculated using the wavefunction $\eta(x,t)$ instead of $\chi(x,t)$,
\textit{i.e.} 
\be
\label{variance_x_free_def}
\Delta x_{\rm undriven}(t) \equiv \sqrt{\left\langle \eta(x,t)\left| \hat{x}^2 \right| \eta(x,t)\right\rangle-\left\langle \eta(x,t)\left| \hat{x} \right| \eta(x,t)\right\rangle^2}\,.
\ee
For the variance of the  momentum we have 
\be
\label{variance_p_1b}
\Delta p(t) \equiv \sqrt{\left\langle \hat{p}^2\right\rangle(t)-\left\langle \hat{p}\right\rangle^2(t)} = \Delta p_{\rm undriven}(t)\,. 
\ee
This means that it remains constant and equal to its initial value at $t=0$.

The solution presented so far,  and  its consequences,  are valid for
any driving function. In the sequel, as a preparation for later Sections, 
we shall focus our attention on  periodic drivings.

\subsection{Floquet approach} 
When $f(t)$ is periodic with period $T$,  
the Schr\"odinger equation (\ref{onebody_schro}) becomes a
differential equation with periodic coefficients where we can
apply  the Floquet theory. This leads us to define the
Floquet Hamiltonian $\hat{H}_F$, which, according to Eq. (\ref{Fl_int}), controls
the time evolution
of the wavefunction at stroboscopic times $t=n T$, with $n\in \mathbb{N}$.
Switching for simplicity to the bra-ket notation, Eq. (\ref{Fl_int}) reads
\be
\label{Floq_Ham_def}
\left|\chi(x,nT)\right\rangle=e^{-i\frac{nT}{\hbar}\hat{H}_F}\,\left|\chi(x,t=0)\right\rangle\,.
\ee
The eigenvalues of the Floquet Hamiltonian will be denoted by $\mathcal{E}_F$
and are known as the quasi-energies. Since $\hat{H}_F$ is hermitian,
they are real numbers. The quasi-energies are the time--like
analogues of the quasi-momenta in the study of crystalline solids. 
Let  $\hat{U}(t,0)=e^{-i \frac{t}{\hbar} \hat{H}}$ be the time evolution operator, i.e. the quantum operator that, when applied to a wavefunction 
describes its evolution from time $0$ to time $t$. According to the Floquet theory and the notation of \cite{Eckardt2017}, we can decompose $\hat{U}(t,0)$ 
as in Eq. (\ref{micro}):
$\hat{U}(t,0) = \hat{U}_F (t,0) \, e^{-i \frac{t}{\hbar} \hat{H}_F}$.
This relation defines the micro-motion operator $\hat{U}_F(t,0)$
in terms of the Floquet Hamiltonian
$\hat{H}_F$ and $\hat{U}(t,0)$. $\hat{U}_F$ is periodic in time
and equals to the unity at every stroboscopic times, implying that
$\hat{U}(nT,0) = e^{-i\frac{nT}{\hbar}\hat{H}_F}$. Therefore
$\hat{U}(t+T,0) = \hat{U}(t,0) \hat{U}(T,0)$. This
means that it is enough to know the evolution operator for times
$t \in [0,T]$ in order to obtain the evolution of the system at all
times $t\ge0$.

The importance of these concepts becomes clear once one realises that
any solution of the time--dependent periodic Schr\"odinger equation
(\ref{onebody_schro}) can be expressed in terms of the Floquet operator and
their eigenfunctions.
Indeed, writing the eigenvalue equation for the
Floquet Hamiltonian
\be
\label{eigeneq_floq_ham}
\hat{H}_F \left | \tilde{u} \right\rangle = \mathcal{E}_F \left | \tilde{u} \right\rangle\,,
\ee
one can apply the micro-motion operator on the wavefunctions
$\left | \tilde{u} \right\rangle$ to write the Floquet modes
(or Floquet functions according to the notation of \cite{Holthaus2016}) as 
\be
\label{floq_modes_def}
\left | u(t) \right\rangle = \hat{U}_F (t,0) \left | \tilde{u} \right\rangle \,,
\ee
which are time--periodic states, as follows from the properties of the
micro-motion operator stated above.
It is now straightforward to construct 
the Floquet states, which are solutions of the time--dependent
Schr\"odinger equation (\ref{onebody_schro}) with periodic $f(t)$:
\be
\label{floq_states_def}
\left | \psi_F(t) \right\rangle = \left | u(t) \right\rangle e^{-i\frac{t}{\hbar}\mathcal{E}_F}\,. 
\ee 
These states form a complete and orthonormal set of eigenfunctions
of the time evolution operator
over a driving period:
\be
\nonumber
\left | \psi_F(t+T) \right\rangle = \hat{U}(t+T,t)\left | \psi_F(t) \right\rangle=e^{-i\frac{T}{\hbar}\mathcal{E}_F}\left | \psi_F(t) \right\rangle\,.
\ee 
Hence, any solution of the Schr\"odinger equation
(\ref{onebody_schro}) can be written as a
superposition of Floquet states as
\be
\label{floq_generic_solution}
 \left | \chi(t) \right\rangle\,=\,\int A(k) \left | u(t) \right\rangle e^{-i\frac{t}{\hbar}\mathcal{E}_F}\,dk \,= \,\int A(k) \left | \psi_F(t) \right\rangle\,dk\,,
\ee
weighted with time--independent coefficients $A$, which depend on the momenta of the particle $k$. Looking at the last expression, notice that the
Floquet states have an occupation probabilities $\left| A \right|^2$
(preserved in time) and a phase factor $e^{-i\frac{t}{\hbar}\mathcal{E}_F}$,
resembling the usual factor $e^{-i\frac{t}{\hbar}E}$
present in any time--evolution of energy eigenstates with eigenvalues $E$
when their Hamiltonian does not depend on time. Therefore the quasi-energies
look as if they were 
effective energies and these are the quantities which determine
the linear phase evolution of the system. Finally, notice that if the system
is prepared in a Floquet state, its time evolution is periodic in time
and in this case it is called a ``quasi-stationary evolution''.

Before obtaining an expression for the micro-motion operator $\hat U_F$
from Eq. (\ref{micro}), it is convenient first to derive an
expression for the Floquet Hamiltonian $\hat{H}_F$ of the system
which will be useful in the many-body case. To get an equation for
$\hat{H}_F$ we need to rewrite Eq. (\ref{complete_sol_1b}) for $t=nT$
in a single exponential operator as in Eq. (\ref{Floq_Ham_def}).
To do this, we can use the Baker-Campbell-Hausdorff formula between
momentum and position exponential operators,  
arriving  at  
\begin{eqnarray}
\label{Floq_Ham_generic_1b}
\hat{H}_F &=  & \frac{\hat{p}^2}{2m}+\left[\frac{\xi(nT)}{nT}+\frac{1}{2m}\int_0^{nT} f(\tau)\,d\tau\right]\hat{p}-\hbar\frac{\theta(x,nT)}{nT}+ \\
& & -\frac{1}{2mnT}\left[\int_0^{nT} f(\tau)\,d\tau\right]\cdot\int_0^{nT} d\tau \left[\int_0^\tau f(\tau')\,d\tau' \right] +\frac{1}{12m}\left[\int_0^{nT} f(\tau)\,d\tau \right]^2\,.\nonumber 
\end{eqnarray}
From this expression one is  tempted to say that the translation of
the center of mass of the wavepacket at different stroboscopic times,
would be $\frac{\xi(nT)}{nT}+\frac{1}{2m}\int_0^{nT} f(\tau)\,d\tau$,
since this is the factor that multiplies the operator $\hat{p}$.
However, this is not true since  to evaluate 
$\left\langle \chi(x,nT) | \hat{x} | \chi(x,nT)\right\rangle$,
one has to split the operators in the exponential recovering 
 the state Eq. (\ref{complete_sol_1b}),
where the translation factor is simply 
$\frac{\xi(nT)}{nT}$. 

Moreover, it is not manifest  from  Eq. (\ref{Floq_Ham_generic_1b}) 
that  the Floquet Hamiltonian is independent of $n$, 
as it should be the case \cite{Eckardt2017}. To clarify this issue
we study in more detail the translational parameter
and the gauge phase. From the first equation in (\ref{vANDdelta_onebody}),
we derive 
\be
\label{transl_parameter}
\xi(t) = -\frac{1}{m} \int_0^t d\tau \left[\int_0^\tau f(\tau')\,d\tau' \right]\,,
\ee
from which follows that
\be
\label{xi_t+T}
\xi(t+T) = \xi(T) + \xi(t) -\frac{t}{m} \int_0^T f(\tau)\,d\tau\,. 
\ee
In a similar way, one gets for the gauge phase:
\be
\label{theta_t+T}
\theta(x,t+T) = \theta(x,T) +\theta(x,t)-\frac{t}{2m\hbar} \left[\int_0^T f(\tau)\,d\tau \right]^2-\frac{1}{m\hbar}\left[\int_0^T f(\tau)\,d\tau\right]\cdot \int_0^t d\tau \left[\int_0^\tau f(\tau')\,d\tau' \right]\,.
\ee
Setting   $t=nT$, with $n\in\mathbb{N}$, 
in the above equations yields
\be
\label{xi_nT}
\xi(nT) = n\xi(T) -\frac{T}{m} \frac{n (n-1)}{2} \int_0^T f(\tau)\,d\tau\,,
\ee
and:
\be
\label{theta_nT}
\theta(x,nT) = n\theta(x,T) -\frac{T}{2m\hbar}\frac{n (n-1)}{2}\left[\int_0^T f(\tau)\,d\tau \right]^2 -\frac{1}{m\hbar}\frac{n(n-1)(2n-1)}{6} \left[\int_0^T f(\tau)\,d\tau\right]\cdot \int_0^T d\tau \left[\int_0^\tau f(\tau')\,d\tau' \right]\,,
\ee
where we used 
\[
\sum_{j=0}^{n-2} (n-j)^2 = \frac{(n-1)(2n^2 +5n+6)}{6}\,.
\]
To continue with the proof 
of the $n$-independence of
the Floquet Hamiltonian, we  split the analysis in two cases:
(1) when the integral of the driving function over one period vanishes, and
(2) when it does not. 

\subsubsection{\bf{$\boldsymbol{\int_0^T f(t) \,dt=0}$}}
When the integral on a time--period is vanishing,
from Eq. (\ref{xi_nT}) we have $\xi(nT)=n\xi(T)$
and therefore the term linear in momentum of the Floquet Hamiltonian
in (\ref{Floq_Ham_generic_1b}) is $n$-independent. Moreover,
since $\xi(nT)$ is linear in terms of the stroboscopic factor $n$,
the stroboscopic motion of the wavepacket has a constant velocity,
as can be inferred from Eq. (\ref{x_mean_1b}).
The constant term in the Floquet Hamiltonian is also trivially
$n$-independent since $\theta(x,nT) =n\theta(x,T) $,
as follows from Eq. (\ref{theta_nT}). Hence, in this case
the Floquet Hamiltonian can be simply written as
\be
\label{Floq_Ham_ft=0_1b}
\hat{H}_F=\frac{\hat{p}^2}{2m}+ \frac{\xi(T)}{T}\hat{p}-\hbar\, \frac{\theta(T)}{T}\,,
\ee
where $\theta(x,T) \equiv \theta(T)$, since the gauge phase is $x$-independent
(in the considered case of: $\int_0^T f(t)\,dt=0$), as one can see from
Eq. (\ref{theta_f(t)}). Moreover, the Floquet Hamiltonian can  be
rewritten as
\be
\nonumber
\hat{H}_F=\frac{\hat{p}^2}{2m}-\frac{\hat{p}}{m} \frac{1}{T} \int_0^T d\tau \int_0^\tau f(\tau')\,d\tau' +\frac{1}{2m}\, \frac{1}{T} \int_0^T d\tau \left[\int_0^\tau f(\tau')\,d\tau' \right]^2\,.
\ee
Notice that we can also express the Hamiltonian in
Eq. (\ref{Floq_Ham_ft=0_1b}) as 
\be 
\hat{H}_F=\frac{[\hat{p}+m\xi(T)/T]^2}{2m} +C\,,
\ee
where $C=-\hbar\theta(T)/T - (m/2) [\xi(T)/T]^2$.
Now,  applying the unitary transformation  
\be 
\hat{\tilde{H}}_F \equiv e^{i a \hat{x}/\hbar} \hat{H}_F e^{-i a \hat{x}/\hbar}\,,
\ee
with $a=m \xi(T)/T$, we get finally 
\be \hat{\tilde{H}}_F=\frac{\hat{p}^2}{2m}+C\,.
\ee
Using these results we can derive  the micro-motion operator
$\hat U_ F$. First of all, from Eq. (\ref{complete_sol_1b}),
the time evolution operator is 
\be
\label{time_evol_op_1b}
\hat{U}(t,0)=e^{i\theta(x,t)} e^{-i\frac{t}{\hbar}\frac{\hat{p}^2}{2m}} e^{-i \frac{\xi(t)}{\hbar} \hat{p}}\,.
\ee
Hence, inverting Eq. (\ref{micro}) and knowing the Floquet
Hamiltonian from Eq. (\ref{Floq_Ham_ft=0_1b}), we get 
\be
\label{micromotion1_ft=0_1b}
\hat{U}_F(t,0) = e^{\frac{i}{\hbar}t \left\{ \left[\frac{\xi(T)}{T}-\frac{\xi(t)}{t}\right]\hat{p}-\hbar\left[\frac{\theta(T)}{T}-\frac{\theta(x,t)}{t}\right] +\frac{1}{2}\left[\int_0^t f(\tau)\,d\tau\right]\cdot\left[\frac{\xi(T)}{T}-\frac{\xi(t)}{t}\right] \right\}}\,,
\ee
where we used the Baker-Campbell-Hausdorff formula.
An alternative expression of the micro-motion operator is 
\be
\label{micromotion2_ft=0_1b}
\hat{U}_F(t,0) = e^{ i t \left[\frac{\theta(x,t)}{t}-\frac{\theta(T)}{T}\right]}\,e^{\frac{i}{\hbar} t \left[\frac{\xi(T)}{T}-\frac{\xi(t)}{t}\right] \hat{p}}\,,
\ee
which has been  derived using the Zassenhaus formula. 

\begin{figure}[t]
\centering
\includegraphics{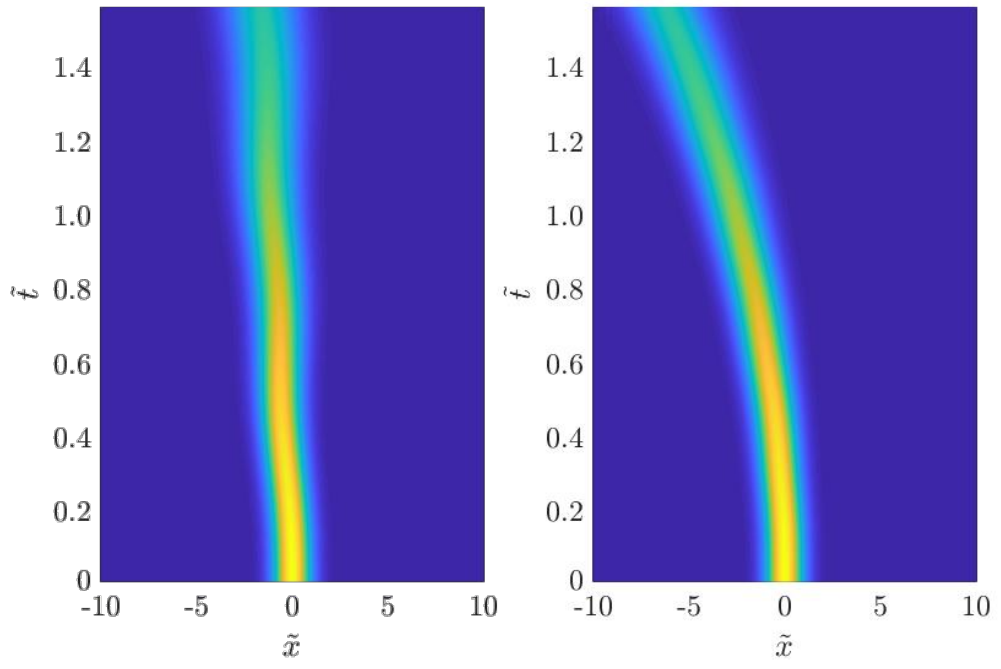}
\includegraphics{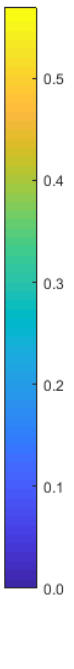}
\caption{Time evolution of density profiles of Gaussian wavepackets
  $\left| \chi(x,t) \right|^2$ for a single particle in a potential:
  $xf(t)$. The left plot shows an evolution with a driving
  force $f(t) = \ell \sin(\omega t)$: the motion proceeds with a
  constant stroboscopic velocity towards  the left. The right plot shows the evolution  under  a driving force
  $f(t) = \ell \sin^2(\omega t)$: the motion is uniformly accelerated
  to negative values of $x$. The figures are
calculated via the split-step Fourier method and
  in both $\tilde{\sigma} = 2^{-1/2}$, $\tilde{\ell}=10$, and
  $\tilde{\omega} = 10$.
   }
\label{fig_1body}
\end{figure}

Let discuss a simple, yet instructive, application of these results.
Imagine we are interested in describing the time evolution of a
Gaussian wavepacket with initial variance $\sigma$ in the infinite
homogeneous space, \textit{i.e.} $\chi(x,0) = \frac{1}{\sqrt[4]{2\pi\sigma^2}} e^{-x^2 / (2\sigma)^2}$. As we saw in the previous Section, in order to
determine its time evolution, we have first to find the eigenvalues and
eigenfunctions of the Floquet Hamiltonian in (\ref{Floq_Ham_ft=0_1b}).
In this case the complete set of eigenfunctions is simply the plane wave set,
and the associated quasi-energies are then easy to determine:
\be
\label{quasien_ft=0_1b}
\left| \tilde{u}\right\rangle = \frac{1}{\sqrt{2\pi}} e^{ikx}\,, \,\,\,\,\,\,\,\,\,\,\,\,\,\, \mathcal{E}_F = \frac{\hbar^2 k^2}{2m} +\frac{\xi(T)}{T} \hbar k -\hbar \frac{\theta(T)}{T}\,,
\ee
where $k$ is the plane wave's momentum. The Floquet modes
can be easily  obtained from the action of $\hat{U}_F$ from
Eq. (\ref{micromotion2_ft=0_1b}) on  the eigenstates
$\left| \tilde{u}\right\rangle$:
\begin{eqnarray} 
\label{floq_modes_ft=0_1b}
\left| u(t)\right\rangle & =  &  \frac{1}{\sqrt{2\pi}} e^{i t \left[\frac{\theta(x,t)}{t}-\frac{\theta(T)}{T}\right] + i k \left[x + \frac{t}{T}\xi(T) -\xi(t)\right]}  \\
&  =  &  \frac{1}{\sqrt{2\pi}} e^{i x \left[k -\frac{1}{\hbar}\int_0^t f(\tau)\,d\tau \right]}\,e^{-i t \left\{\frac{1}{2m\hbar t}\left[\int_0^t d\tau \left(\int_0^\tau f(\tau')\,d\tau' \right)^2 -\frac{t}{T}\int_0^T d\tau \left(\int_0^\tau f(\tau')\,d\tau' \right)^2\right] +k\left[\frac{\xi(t)}{t}-\frac{\xi(T)}{T}\right]\right\}}\,,
\nonumber 
\end{eqnarray}
where in the second equality we used Eq. (\ref{theta_f(t)}).
The Floquet modes are plane waves with a momentum that varies in time, 
\[
\left\langle u(t) \left|\hat{k}\right| u(t) \right\rangle \,=\, k-\frac{1}{\hbar}\int_0^t f(\tau)\,d\tau \,,
\]
and which return to their  initial value  $k$ at stroboscopic times.
As required,  the Floquet modes are time--periodic with period $T$. 
The Floquet states are obtained from Eqs. (\ref{floq_states_def})
and (\ref{quasien_ft=0_1b}), 
\be
\label{floq_states_ft=0_1b}
\left| \psi_F(t) \right\rangle = \frac{1}{\sqrt{2\pi}} e^{i\left[k x +\theta(x,t)\right]-it\frac{\hbar k^2}{2m} - i k\xi(t)}\,.
\ee
They are plane waves, periodic in time with period $T$ and their
momentum expectation value varies in the same way as does for
the Floquet modes. One can now evaluate the time evolution
of the Gaussian wavepacket from Eq. (\ref{floq_generic_solution}).
In order to do so, we  compute the amplitude $A(k)$ 
\be
\nonumber
A(k) = \int_{-\infty}^\infty \chi(x,0) \psi_F^* (x,0) = \sqrt[4]{\frac{2\sigma^2}{\pi}} e^{-(k\sigma)^2}\,, 
\ee
and perform  the Gaussian integration in
Eq. (\ref{floq_generic_solution}), arriving at 
\be
\label{GWP_timeevolved}
\chi(x,t)\,=\,\frac{1}{\sqrt[4]{2\,\pi\,\sigma^2}}\,\frac{e^{i\,\theta(x,t)}}{\sqrt{1+i\,\frac{\hbar\,t}{2\,m\,\sigma^2}}}e^{-\frac{\left[x-\xi(t)\right]^2}{4\left(\sigma^2+i\,\frac{\hbar\,t}{2\,m}\right)}}\,.
\ee
The wavepacket has  a Gaussian shape centered at  $\xi(t)$ and spreads  in time as 
\be
\label{GWP_variance}
\Delta x(t)\,=\,\sqrt{\sigma^2+\frac{\hbar^2\,t^2}{4\,m^2\,\sigma^2}}\,,
\ee
in agreement with  Eq. (\ref{variance_x_1b}).
The left side of Fig. \ref{fig_1body} shows an  example,
where $f(t) = \ell \sin(\omega t)$. The center of mass of the wavepacket
is located at 
$\xi(t) = \frac{\ell}{m\omega^2} \left[\sin(\omega t)-\omega t\right]$,
and it spreads according to Eq. (\ref{GWP_variance}).
We use  the parameterization 
$\ell = l \cdot \tilde{\ell}$ and $\omega = u \cdot\tilde{\omega}$,
where $\tilde{\ell}$ and $\tilde{\omega}$ are dimensionless,  and define $\tilde{t} = t/u$ and
$\tilde{x} = x \sqrt[3]{\frac{m l}{\hbar^2}}$. In the left
side of Fig. \ref{fig_1body} we set: $\tilde{\sigma} = \sigma \sqrt[3]{\frac{m l}{\hbar^2}} = 2^{-1/2}$, $\tilde{\ell}=10$, and $\tilde{\omega} = 10$.

\subsubsection{\bf{$\boldsymbol{\int_0^T f(t) \,dt \neq 0}$}}
In this case, the independence of the Floquet Hamiltonian  (\ref{Floq_Ham_generic_1b})
on $n$ is more difficult to demonstrate. 
Let us define a function $F(t)$, such that $\frac{dF}{dt} = f(t)$. 
We have $\int_0^T f(t)\,dt = F(T) = c$, where $c$ 
depends on the driving parameters and,  by definition,  $F(0) = 0$. 
It follows that  $F(nT) = n F(T) = n c$. It is  easy
to prove that $F(t+T) = F(T) + F(t) = c+F(t)$.
Therefore $F(nT+t) = nc+F(t)$ and $\xi(nT)$ can  be written as
\be
\label{xi_ft!=0}
\xi(nT) = -\frac{1}{m}\int_0^{nT} F(t)\,dt = -\frac{n^2}{m} I\,,
\ee
where $I = \int_0^T F(t)\,dt$. Thus  $\xi(nT)$
depends quadratically on the stroboscopic factor $n$, and 
the stroboscopic motion experiences  a uniform 
acceleration $-\frac{1}{m} I$. Next, since $\xi(nT) \propto n^2$, one has $\xi(-T) = \xi(T)$ and, choosing $n=-1$ in Eq. (\ref{xi_nT}),
yields: $\xi(T) = -\frac{T}{2m}\int_0^T f(t)\,dt$. This can be substituted
back into Eq. (\ref{xi_nT}) to obtain
\be
\label{xi_nT_ft!=0}
\xi(nT) = -\frac{n^2 T}{2m}\int_0^T f(t)\,dt\,.
\ee 
If we now take $t=nT$ in (\ref{transl_parameter}) and use \eqref{xi_nT_ft!=0},  we derive the relevant equation
\be
\nonumber
\int_0^{nT} d\tau \int_0^\tau f(\tau')\,d\tau' \,=\, \frac{n^2 T}{2} \int_0^T f(t)\,dt\,, 
\ee
that holds when the  integral of the driving function over a driving period does not vanish. 
Using these results into  (\ref{Floq_Ham_generic_1b}), we can write
\be
\label{Floq_Ham_ft!=0_1b}
\hat{H}_F = \frac{\hat{p}^2}{2m} -\hbar\frac{\theta(x,T)}{T}-\frac{1}{6m} \left[\int_0^T f(\tau)\,d\tau\right]^2\,,
\ee
or, equivalently, 
\be
\nonumber
\hat{H}_F = \frac{\hat{p}^2}{2m}+x\,\frac{1}{T} \int_0^T f(\tau)\,d\tau+\frac{1}{2m}\frac{1}{T} \int_0^T d\tau \left[\int_0^\tau f(\tau')\,d\tau' \right]^2-\frac{1}{6m} \left[\int_0^T f(\tau)\,d\tau\right]^2\,.
\ee

This expression is independent on $n$, a fact which completes the proof. 
Unlike the case where  $\int_0^T f(t) \,dt =0$, 
the Floquet Hamiltonian does not contain a term proportional to  $\hat{p}$,
but  a static linear potential. This term forces the particle to move
to the left/right for positive/negative  values of $\int_0^T f(t) \,dt$. 
An example is given in   Fig. \ref{fig_1body}-right where  
$\frac{1}{T}\int_0^T f(\tau)\,d\tau = \frac{\ell T}{2} >0$, so that  
the wavepacket moves with an acceleration of $-\frac{\ell T^2}{4m}$. 
However,  its spread does not depends on the external driving force
as  predicted in Eq. (\ref{variance_x_1b}).

The eigenfunctions of the Floquet Hamiltonian are the Airy function $Ai$
\cite{LandauLif} of the form:
\be
\label{utilde_ft!=0_1b}
\tilde{u}(x) = C Ai\left\{\left(\frac{2mT^2}{\hbar^2 \left[\int_0^T f(\tau)\,d\tau\right]^2}\right)^{1/3} \, \left( \frac{x}{T}\int_0^T f(\tau)\,d\tau - \mathcal{E}_F + \Omega \right)\right\}\,,
\ee
where $C$ is a normalization constant and
\[
\Omega = \frac{1}{2m}\frac{1}{T} \int_0^T d\tau \left[\int_0^\tau f(\tau')\,d\tau' \right]^2-\frac{1}{6m} \left[\int_0^T f(\tau)\,d\tau\right]^2\,.
\] 
The Floquet Hamiltonian has a continuous spectrum spanning
the whole range of energy values $\mathcal{E}_F$ from
$-\infty$ to $+\infty$.

The micro-motion operator is obtained inverting Eq. (\ref{micro}), and it leads to 
\be
\label{micromotion_ft!=0_1b}
\hat{U}_F(t,0) = e^{\frac{i}{\hbar} \left\{ t\hbar \left[\frac{\theta(x,t)}{t}-\frac{\theta(x,T)}{T}\right] - \frac{t}{2m} \int_0^T f(\tau)\,d\tau \cdot \left[\frac{1}{3}\left(1+2\frac{t^2}{T^2}\right)\int_0^T f(\tau)\,d\tau \right]-\xi(t)\frac{t}{T}\int_0^T f(\tau)\,d\tau \right\}}\, e^{-\frac{i}{\hbar} \left[\xi(t) + \frac{t^2}{2mT} \int_0^T f(\tau)\,d\tau\right] \hat{p}}\,.
\ee
This expression makes it complicated to determine the time evolution, even for a Gaussian wavepacket,
using Eq.(\ref{floq_generic_solution}). 
To circumvent this problem we perform 
the  unitary transformation 
\be
\nonumber
\chi(x,t) = \hat{U}_F(t,0) \tilde{\chi}(x,t)\,, 
\ee
where the transformed wavefunction
satisfies 
\cite{Holthaus2016}
\be
\nonumber
i\hbar \frac{\p \tilde{\chi}}{\p t} = \hat{H}_F \tilde{\chi}(x,t)\,.
\ee
Since $\hat{H}_F$ has a linear potential term, we can apply the
same reasoning used to solve the original equation (\ref{onebody_schro})
for a constant driving function
$\tilde{f}=\frac{1}{T} \int_0^T f(\tau)\,d\tau$, therefore
we translate and gauge transform the wavefunction
$\tilde{\chi}(x,t)$ in order to wash out the $x$-linear term
in the Floquet Hamiltonian. By doing so, we finally get
Eq. (\ref{complete_sol_1b}), which is
thus the convenient way to obtain the time--evolved wavepacket.
In summary, we need first to calculate the free expansion of $\chi(x,0)$, then
to translate the solution and finally to multiply it by the gauge phase.

The detailed analysis performed so far is valid for a single particle
subjected to a linear potential which varies periodically in time. 
We shall show below that it can be extended straightforwardly 
to  two- or many-particles interacting with
a generic interacting potential $V_{2b}(x_j-x_i)$.

\section{Introducing interactions: The two-body problem}
Let us now consider a one--dimensional system of two interacting particles
subjected to a linear time--periodic potential. The Schr\"odinger equation
reads
\be
\label{Schro_twobodies_generic}
i\,\hbar\,\frac{\p \chi}{\p t}=\sum_{j=1}^2\left[-\frac{\hbar^2}{2\,m} \frac{\p^2}{\p x_j^2}+x_j \,f(t)\right]\chi+V_{2b}(x_2-x_1) \chi\,,
\ee
where $V_{2b}(x_2-x_1)$ is a generic 
potential between the two particles. To solve the Schr\"odinger equation
(\ref{Schro_twobodies_generic}), we can employ the same method discussed
in the previous Section: First we perform the gauge transformation 
\be
\label{wavefunction_gaugetrasf_twobodies}
\chi (x_1,x_2,t) = \, e^{i \left[\theta(x_1,t)+\theta(x_2,t)\right]} \eta(y_1(t), y_2(t), t)\,,
\ee
where $y_j(t)=x_j-\xi(t)$, for $j=1,2$.
The wavefunction $\eta(y_1,y_2,t)$ satisfies the Schr\"odinger equation
for two interacting particles with no external potential:
\be
\label{twobodies_schro_eta}
i \hbar \frac{\p \eta}{\p t}= -\frac{\hbar^2}{2\,m} \left[\frac{\p^2 }{\p y_1^2}+\frac{\p^2 }{\p y_2^2}\right]\eta + V_{2b}(y_2 -y_1)\,\eta\,,
\ee
while $\xi(t)$ and $\theta(x_j,t)$ obey Eqs. (\ref{transl_parameter})
and (\ref{theta_f(t)}), once we 
use the same initial conditions of
the previous case.

Notice that $V_{2b}(y_1 -y_2)\,=\,V_{2b}(x_1 -x_2)$,
because   $y_j(t)=x_j-\xi(t)$. Moreover, since $\xi(0)=0$, the two wavefunctions coincide at initial time:
$\chi(x_1,x_2,0)\,=\,\eta(x_1,x_2,0)$, hence the solution of \eqref{Schro_twobodies_generic}
can be written as
\be
\label{complete_sol_2b}
\chi(x_1,x_2,t)\,=\,e^{i\theta(x_1,t)+i\theta(x_2,t)} e^{-i \frac{\xi(t)}{\hbar} (\hat{p}_1+\hat{p}_2)} e^{-i\frac{t}{\hbar}\left[\frac{\hat{p}_1^2+\hat{p}_2^2}{2m} +V_{2b}(x_2-x_1)\right]}\,\chi(x_1, x_2,0)\,.
\ee
With this expression, using the procedure discussed in the previous Section,
we can compute the expectation values of physical observables and their variances. More precisely,
the expectation value of a single particle operator $\hat{O}_j$ is defined as 
\be
\label{expectation_value_def}
\left\langle \hat{O}_j \right \rangle(t) \equiv \left\langle \chi(x_1,x_2,t)\left| \hat{O}_j \right| \chi(x_1,x_2,t)\right\rangle = \int_{-\infty}^\infty dx_1 \int_{-\infty}^\infty dx_2\, \chi^*(x_1,x_2,t)\, \hat{O}_j \,\chi(x_1,x_2,t)\,,
\ee
and expectation values of position and momentum can be computed
using the Baker-Campbell-Hausdorff formula.

We will show below  that there is a decoupling
between the linear potential term and the interacting one. 
This decoupling arises from the separation of 
the center of mass motion (which is determined by the
external potential), and the relative motion (determined by 
the interacting potential).
The diffusion  of the wavepacket evolves as it would be free from the
linear time dependent potential, but of course depends on the 
interaction. 

The undriven Hamiltonian is given by 
$$\hat{H}_0 = \frac{\hat{p}_1^2+\hat{p}_2^2}{2m} +V_{2b}(x_2-x_1)\,.$$ This implies that the 
total momentum of the system
$\hat{P} = \hat{p}_1+\hat{p}_2$ is conserved, i.e. $\left[\hat{H}_0,\hat{P}\right]=0$. An example is 
the contact interaction
$V_{2b}(x_2-x_1)\,=\,\lambda\,\delta(x_2-x_1)$, with $\lambda$ 
the coupling strength. This property allows us to  calculate the total energy of the state:
\be
\label{Enery_def}
E(t) = \left\langle \hat{H}\right\rangle (t) = \left\langle \chi(x_1,x_2,t)\left| \left[\frac{\hat{p}_1^2 + \hat{p}_2^2}{2m}+ f(t) \left(x_1 + x_2 \right) +V_{2b}\left(x_2-x_1\right) \right]\right| \chi(x_1,x_2,t)\right\rangle\,. 
\ee
After a lengthy calculation, using the canonical commutation
relations and Eq. (\ref{complete_sol_2b}), we obtain  for a generic driving function $f(t)$, including as well the non-periodic cases: 
\begin{eqnarray}
\label{energy_result_2b}
E(t) &=& E(0) +\frac{1}{m}\left[\int_0^t f(\tau)d\tau \right]^2+\sum_{j=1}^2 p_{0,j} \left[\frac{t}{m}f(t) -\frac{1}{m}\int_0^t f(\tau)d\tau \right] +\\
&& - \frac{2 f(t)}{m}\int_0^t d\tau \int_0^\tau f(\tau')d\tau' +\sum_{j=1}^2 x_{0,j} \left[f(t)-f(0)\right]\,,\nonumber
\end{eqnarray}
where $E(0)$ is the initial energy of the state, containing 
all the interaction effects.
The remaining  terms arise
from the linear driving potential and depend on the 
position $x_{0,j}$ and momenta
$p_{0,j}$, of the $j$-th particle at time $t=0$. 
If $f(t)$ is constant, as 
for a constant (gravitational or electric) force,
then the energy is conserved. On the other hand, 
if $f(t)$  is periodic, its integral over a 
time--period vanishes, and  $f(t=0)=0$, then  the energy is conserved at
stroboscopic times.

Next we shall  study  the models where $f(t)$ is periodic.
As done in the previous Section,  we shall consider two cases: 
$\int_0^T f(t)\,dt =0$, and $\int_0^T f(t)\,dt\neq0$.
The evolution operator can be read from \eqref{complete_sol_2b}:
\be
\label{time_evolop_2b}
\hat{U}(t,0) \,=\, e^{i\left[\theta(x_1,t)+\theta(x_2,t)\right]} e^{-i \frac{\xi(t)}{\hbar} (\hat{p}_1+\hat{p}_2)} e^{-i\frac{t}{\hbar}\left[\frac{\hat{p}_1^2+\hat{p}_2^2}{2m} +V_{2b}(x_2-x_1)\right]}\, . 
\ee

It is convenient to use the center of mass and relative coordinates:
$x=x_2-x_1$ and $X=\frac{x_1+x_2}{2}$.  In these variables the effects
of the linear time dependent potential and the  interactions
are completely decoupled. The time evolution in these coordinates reads 
\be
\label{time_evolvop_2b_decoupled}
\hat{U}(t,0) = \hat{U}^{\rm com}(t,0) \hat{U}^{\rm rel}(t,0) = e^{-\frac{i}{\hbar}\left\{2X\int_0^t f(\tau)\,d\tau +\frac{1}{m}\int_0^t d\tau \left[\int_0^\tau f(\tau')\,d\tau'\right]^2\right\}} e^{-i \frac{\xi(t)}{\hbar} \hat{P}} e^{-i\frac{t}{\hbar} \frac{\hat{P}^2}{4m}} e^{-i\frac{t}{\hbar}\left[\frac{\hat{p}^2}{m} +V_{2b}(x)\right]}\,,
\ee
where ${\hat P}$ is the total momentum, that commutes with the undriven Hamiltonian, and 
$\hat{p}=\hat{p}_2 -\hat{p}_1$, is the relative momentum of the particles.

\subsubsection{\bf{$\boldsymbol{\int_0^T f(t) \,dt=0}$}}
In this case one finds 
\be
\label{Floq_HamG_twobodies1}
\hat{H}_F=\sum_{j=1}^2\left[\frac{\hat{p}_j^2}{2\,m}+\frac{\xi(T)}{T}\hat{p}_j-\hbar\frac{\theta(T) }{T}\right]+V_{2b}(x_2-x_1)\,,
\ee
where $\theta(x_j,T) = \theta(T)$,
as follows  from Eq. (\ref{theta_f(t)}).

From the analysis performed so far, and for the similarities
with the one-body case, we know that the stroboscopic
motion described by the Floquet Hamiltonian occurs with a constant velocity,
since the translational parameter is: $\xi(nT) \propto n$. Notice that
if the Schr\"odinger equation with the original undriven Hamiltonian is
solvable, then also the Floquet Hamiltonian associated to the
motion under the action of a linear time dependent potential is solvable,
since it is described by the same two-body potential of the original
problem with no driving, apart from a momentum shift.
We observe that it is not convenient to solve the dynamics via
Eq. \eqref{floq_generic_solution} with respect to the eigenfunctions
of the Floquet Hamiltonian in Eq.  (\ref{Floq_HamG_twobodies1}),  while it
is instead more advantageous to pass to relative and center of mass
coordinates. 
Using the center of mass and relative coordinates the Floquet 
 Floquet Hamiltonian decouples in two parts

\be
\label{floq_ham_com_ft=0}
\hat{H}_F^{\rm com} = \frac{\hat{P}^2}{4m}+\frac{\xi(T)}{T}\hat{P} -2\hbar\frac{\theta(T)}{T} \,,
\ee
and
\be
\label{floq_ham_rel_ft=0}
\hat{H}_F^{\rm rel} = \frac{\hat{p}^2}{m}+V_{2b}(x)\,.
\ee
The same factorization occurs for the  micro-motion operators, by defining
\be
\hat{U}(t,0) = \hat{U}_F^{\rm com}(t,0) e^{-i\frac{t}{\hbar} \hat{H}_F^{\rm com}}\,\hat{U}_F^{\rm rel}(t,0) e^{-i\frac{t}{\hbar} \hat{H}_F^{\rm rel}}\,.
\ee
Using Eq. (\ref{time_evolvop_2b_decoupled}), the micro-motion operator for  the center of mass evolution has a form
\be
\label{micromotion_com_ft=0}
\hat{U}_F^{\rm com}(t,0) = e^{-i t \left\{\frac{2X}{\hbar t}\int_0^t f(\tau)\,d\tau +\frac{1}{m\hbar t}\int_0^t d\tau \left[\int_0^\tau f(\tau')\,d\tau'\right]^2 +2\frac{\theta(T)}{T}\right\}} \,e^{i\frac{t}{\hbar} \left[\frac{\xi(T)}{T}-\frac{\xi(t)}{t}\right] \hat{P}}\,,
\ee
while 
the micro-motion operator for the relative coordinate is instead trivial, 
\be
\label{micromotion_rel_ft=0}
\hat{U}_F^{\rm rel}(t,0) =\hat{\mathbbm{1}}\,.
\ee
The time evolution for the relative motion depends of course on 
 the interacting potential $V_{2b}(x)$. 
 Concerning the center of mass motion, we notice  the
similarity of Eq. (\ref{floq_ham_com_ft=0}) with the Floquet
Hamiltonian (\ref{Floq_Ham_ft=0_1b})
for a single particle, that allow us to use the results of the previous Section. 
The eigenfunctions of the Floquet Hamiltonian (\ref{floq_ham_com_ft=0})
are plane waves with a continuous spectrum of quasi-energies:

\be
\label{quasien_ft=0_2b}
\tilde{u}^{\rm com}(X)  = 
\frac{1}{\sqrt{2\pi}} e^{iKX}\,, \,\,\,\,\,\,\,\,\,\,\,\,\,\, \mathcal{E}_F^{\rm com} = \frac{\hbar^2 K^2}{4m} +\frac{\xi(T)}{T} \hbar K -2\hbar \frac{\theta(T)}{T}\,,
\ee
where $K$ is the center of mass momentum.
Next, we can get the Floquet modes by applying $\hat{U}_F^{\rm com}(t,0)$
onto  $\tilde{u}^{\rm com}(X)$, obtaining 
\be
\label{floq_modes_ft=0_2b}
{u}^{\rm com}(X,t)  = \frac{1}{\sqrt{2\pi}} e^{i X \left[K -\frac{2}{\hbar}\int_0^t f(\tau)\,d\tau \right]}\,e^{-i t \left\{\frac{1}{m\hbar t}\left[\int_0^t d\tau \left(\int_0^\tau f(\tau')\,d\tau' \right)^2 -\frac{2t}{T}\int_0^T d\tau \left(\int_0^\tau f(\tau')\,d\tau' \right)^2\right] +K\left[\frac{\xi(t)}{t}-\frac{\xi(T)}{T}\right]\right\}}\,,
\ee
where we used Eq. (\ref{theta_f(t)}).
As in the one-body problem, the Floquet modes
are plane waves with a momentum varying  in time as 
\[\left\langle u(t) \left|\hat{K}\right| u(t) \right\rangle \,=\, K-\frac{2}{\hbar}\int_0^t f(\tau)\,d\tau\,,
\]
which implies that $\left\langle K\right\rangle(nT)=K$.
We finally get the Floquet states from Eq. (\ref{floq_states_def})
and (\ref{quasien_ft=0_2b}), 
\be
\label{floq_states_ft=0_2b}
\psi_F^{\rm com}(X,t) =
 \frac{1}{\sqrt{2\pi}} e^{i\left\{K X -\frac{2X}{\hbar t}\int_0^t f(\tau)\,d\tau -\frac{1}{m\hbar t}\int_0^t d\tau \left[\int_0^\tau f(\tau')\,d\tau'\right]^2 \right\}-it\frac{\hbar K^2}{4m} - i K\xi(t)}\,, 
\ee
that  are plane waves, periodic in time with period $T$,
and whose  average center of mass momentum behaves like that 
of  the Floquet modes. Therefore the center of mass component
of the wavefunction, solution of  (\ref{Schro_twobodies_generic}), reads as 
\be
\label{com_wavefunction_ft=0}
\phi(X,t) = \int A(K) \psi_F^{\rm com}(X,t)\,dK\,,
\ee
where we have written: $\chi(x_1,x_2,t) = \phi(X,t) \varphi(x,t)$.

\subsubsection{\bf{$\boldsymbol{\int_0^T f(t) \,dt \neq 0}$}}
Using the methods presented  in previous Sections,
we find 
\be
\label{Floq_HamG_twobodies2}
\hat{H}_F = \sum_{j=1}^2 \left[\frac{\hat{p}_j^2}{2m} -\hbar \frac{\theta(x_j,T)}{T}\right]-\frac{1}{3m}\left[\int_0^T f(\tau)\,d\tau \right]^2 + V_{2b}(x_2 -x_1)\,. 
\ee
This expression contains  a linear potential, hidden in the gauge phases $\theta(x_j,T)$.
Analogously to the
one-body example,
the stroboscopic motion of the particles
is uniformly accelerated, 
\[
\frac{d^2\left\langle x_j \right\rangle}{dt^2}(nT)\,=\, -\frac{1}{m}\int_0^T d\tau \int_0^\tau f(\tau')\,d\tau' \,.
\] 
Using the center of mass and relative coordinates, the Floquet Hamiltonian
(\ref{Floq_HamG_twobodies2}) splits in two parts 
\be
\label{floq_ham_com_ft!=0}
\hat{H}_F^{\rm com} = \frac{\hat{P}^2}{4m}+\hat{X} \frac{1}{T}\int_0^T f(\tau)\,d\tau +\frac{1}{m}\frac{1}{T}\int_0^T d\tau \left[\int_0^\tau f(\tau')\,d\tau'\right]^2 -\frac{1}{3m}\left[\int_0^T f(\tau)\,d\tau \right]^2\,,
\ee
while the Floquet Hamiltonian of the relative motion is given by 
Eq. (\ref{floq_ham_rel_ft=0}). The difference between
the  cases  (1) and (2) 
stems only from the center of
mass motion which has an additional linear dependence on $\hat{P}$ in the first case,  and $\hat{X}$ in the second. 
The micro-motion operator can be split as well, obtaining 
Eq. (\ref{micromotion_rel_ft=0}) for the relative part, and
\begin{multline}
\label{micromotion_com_ft!=0}
\hat{U}_F^{\rm com}(t,0) = e^{\frac{i}{\hbar} \big\{ t \left[X\left(\frac{1}{T}\int_0^T f(\tau)\,d\tau -\frac{1}{t}\int_0^t f(\tau)\,d\tau \right) +\frac{1}{mT} \int_0^T d\tau \left[\int_0^\tau f(\tau')\,d\tau'\right]^2 +\frac{1}{mt}\int_0^t d\tau \left[\int_0^\tau f(\tau')\,d\tau'\right]^2 \right] +}\\
^{- \frac{t}{m} \int_0^T f(\tau)\,d\tau \cdot \left[\frac{1}{3}\left(1+2\frac{t^2}{T^2}\right)\int_0^T f(\tau)\,d\tau \right]+-2\xi(t)\frac{t}{T}\int_0^T f(\tau)\,d\tau \big\}}\, e^{-\frac{i}{\hbar} \hat{P} \left[\xi(t) + \frac{t^2}{2mT} \int_0^T f(\tau)\,d\tau\right]}\,,
\end{multline}
for the center of mass.

The dynamics of the relative part can be analysed once the
two-body potential is given, while the analysis performed on the center
of mass part follows the same line of the one-body case.
By this we mean that one has to perform a
unitary transformation on the center of mass wavefunction:
$\Phi(X,t) = \hat{U}_F \tilde{\Phi}(X,t)$, and therefore
the new wavefunction $\tilde{\Phi}(X,t)$ satisfies a time dependent
Schr\"odinger equation with the Floquet Hamiltonian
(\ref{floq_ham_com_ft!=0}). Washing away the $X$-linear dependence
of the Floquet Hamiltonian by means of a translation and a
gauge transformation, for the center of mass part of
Eq. (\ref{complete_sol_2b}) we have 
\be
\label{com_wavefunction_evol}
\Phi(X,t) = e^{-\frac{i}{\hbar}\left\{2X\int_0^t f(\tau)\,d\tau +\frac{1}{m}\int_0^t d\tau \left[\int_0^\tau f(\tau')\,d\tau'\right]^2\right\}} e^{-i \frac{\xi(t)}{\hbar} \hat{P}} e^{-i\frac{t}{\hbar} \frac{\hat{P}^2}{4m}}\, \Phi(X,0)\,,
\ee
where Eq. (\ref{time_evolvop_2b_decoupled}) has been used.

As an example, we use the above results to study the time evolution of two particles with contact interactions initially prepared in a Gaussian wavepacket. 

\subsection{Contact interactions}
Let consider  a contact potential:
$V_{2b}(x_2-x_1) = \lambda \delta(x_2-x_1)$, where
$\lambda>0$ is the repulsive interaction parameter. At the initial time we 
 prepare a Gaussian wavepacket with variance $\sigma$
\be
\label{initial_WP_2b}
\chi(x_1,x_2,0) = \frac{1}{\sqrt{\pi \sigma^2}} e^{-\left(x_1^2+x_2^2\right)/2\sigma^2}\,, 
\ee
that factorizes into  the center of mass and relative parts 
\be
\label{com_rel_wavepacket}
\Phi(X,0) = \sqrt[4]{\frac{2}{\pi\sigma^2}} e^{-X^2 /\sigma^2}\,\,\,\,\,\, ,\,\,\,\,\,\,\,\,\,\,\,\,\,\,\,\,\,\varphi(x,0)=\frac{1}{\sqrt[4]{2\pi\sigma^2}}e^{-x^2/4\sigma^2}\,.
\ee

Let us start with the case: $\int_0^T f(\tau)\,d\tau =0$.
Finding the time--independent coefficient $A(K)$ appearing in  
Eq. (\ref{com_wavefunction_ft=0}) at  $t=0$, and using
(\ref{com_rel_wavepacket}), yields:
\be
\label{com_evolved}
\Phi(X,t)=\sqrt[4]{\frac{2}{\pi \sigma^2}}\,\frac{e^{i\theta(X,t)}}{\sqrt{1+i\frac{\hbar \,t}{m \sigma^2}}}\, \,e^{-\frac{\left[X-\xi(t)\right]^2}{\sigma^2 \left(1+i\frac{\hbar\,t}{m\sigma^2}\right)}}\,.
\ee

\begin{figure}[t]
\centering
\includegraphics[width=151mm]{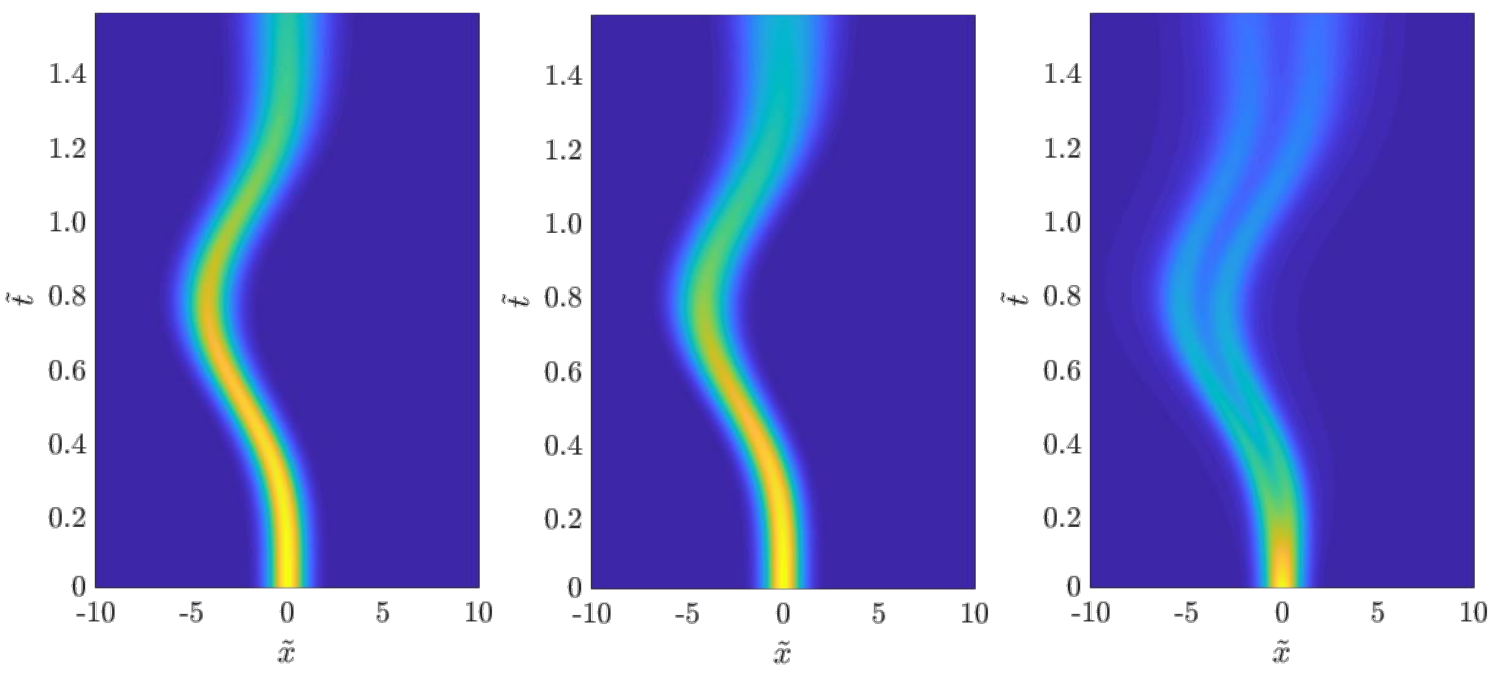}
\includegraphics[width=7mm]{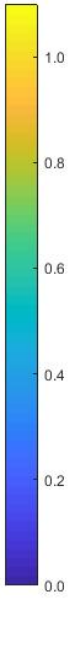}
\caption{Time evolution of density matrix profiles (\ref{density_matrix_def}) for a Gaussian
  wavepacket (\ref{initial_WP_2b}), under the action of a linear
  external potential: $x f(t)$, with driving function
  $f(t) = \ell \left[\cos^2(\omega t) -1 + \frac{4}{3} \sin^4(\omega t)\right]$. The left side plot is the free case, $\tilde{\lambda}=0$,
  the central plot has $\tilde{\lambda}=1$, while the right side plot has $\tilde{\lambda}=\infty$. The center of mass moves with constant
  stroboscopic velocity, as predicted analytically, and the wavepacket
  spreads over time as it would do for the undriven case $\ell=0$. As one can see from the right side plot,
  for very large interactions, the wavepacket rapidly tends to split 
  in two specular parts. In all the figures the values
  $\tilde{\ell}=200$, $\tilde{\omega}=2$ and $\tilde{\sigma}=1$ have been
  chosen.}
\label{2b_ft_equal0}
\end{figure}

Concerning  the relative motion, we use  the propagator $G(x,x';t,0)$ in the presence of a  Dirac 
$\delta$-potential  \cite{Bauch1985,Andreata2004}
\be
\label{propagator_eq}
\varphi(x,t)=\int_{-\infty}^{\infty} G(x,x';t,0)\,\varphi(x',0)\,dx'\,,
\ee
with 
\be
\label{propagator}
G(x,x';t,0)=\frac{1}{\sqrt{4\,\pi\,i\,\hbar\,t/m}}\,e^{i\,\frac{m\,(x-x')^2}{4\,\hbar\,t}}-\frac{m\,\lambda}{4\,\hbar^2}\,e^{\frac{m\,\lambda}{2\,\hbar^2}\left(\left|x\right| +\left|x'\right|\right) +i\,\frac{m\,\lambda^2\,t}{4\,\hbar}}\,{\rm erfc}\left(\frac{\left| x\right|+\left| x'\right|+i\frac{\lambda\,t}{\hbar}}{\sqrt{4\,i\,\hbar\,t/m}}\right)\,,
\ee
with ${\rm erfc}$ being the complementary error function:
\be
\nonumber
{\rm erfc}(z)=\frac{2}{\sqrt{\pi}}\int_z^\infty e^{-t^2}\,dt\,.
\ee
The  numerical  integration  of 
(\ref{propagator_eq}), provides the wavefunction $\chi(x_1,x_2,t)$ for any value of $\lambda >0$. 
In the limit of  hard--core interactions, 
$\lambda \to \infty$, the integral (\ref{propagator_eq}) can be computed analytically 
\be
\label{relative_twobody_TG}
\varphi(x,t)=\frac{1}{\left(2\pi\right)^{1/4}}\sqrt{\frac{i m \sigma/\hbar t}{-1+i m \sigma^2/\hbar t}} \,\,{\rm erf}\left(\frac{m\,\sigma\,x}{2\hbar t \sqrt{-1+i m\sigma^2/\hbar t}} \right)\,e^{-\frac{m}{4\hbar t}\frac{x^2}{i+m\sigma^2/\hbar t}}\,,
\ee
where ${\rm erf}(z)=1-{\rm erfc}(z)$. We have
studied the time evolution of the density matrix 
\be
\label{density_matrix_def}
\rho(x_1,t)=2\,\int_{-\infty}^\infty \left| \chi(x_1,x_2,t)\right|^2\,dx_2\,,
\ee
in order to visualize the evolution of the wavepacket. The density
matrix (\ref{density_matrix_def}) reads in 
 the center of mass and relative wavefunctions, as 
\be
\label{density_matrix_calc}
\rho(x,t)=2\,\int_{-\infty}^\infty \left| \Phi\left(\frac{x_1}{2}+x,t\right)\right|^2\,\left|\varphi(x_1,t) \right|^2\,dx_1\,,
\ee
The results are reported in Fig. \ref{2b_ft_equal0}
for different times and coupling strengths $\lambda$,
using the driving function 
\[
f(t) = \ell \left[\cos^2(\omega t) -1 + \frac{4}{3} \sin^4(\omega t)\right]\,.
\]
We choose  the same dimensionless variables as in 
the one-body case: dimensionless coupling
strength  $\tilde{\lambda} = l \frac{m\lambda}{\hbar^2}$, 
 $\tilde{\ell}=200$, $\tilde{\omega}=2$ and $\tilde{\sigma}=1$. 
 The values $\tilde{\lambda} = 0$, $1$ and $\infty$, correspond to the left,
center and right sides of  Fig. \ref{2b_ft_equal0}. Here 
\[
\xi(t) = \frac{\ell}{12 m\omega^2} \sin^4(\omega t)\,,
\] 
vanishes at stroboscopic times, as checked 
in  the numerical simulations. We have also verified
that the wave packet expands as it were not subjected to the linear
oscillating potential, in agreement with the theoretical prediction. 

Fig. \ref{2b_ft_equal0} shows that increasing
the  parameter  $\lambda$, the variance of the wavepacket increases
in time more rapidly. We  have been able to  fit this behaviour with the  approximation 
\be
\label{variance_interactions}
\Delta x_j (t) \approx \frac{\sigma}{\sqrt{2}} \sqrt{1+\left(\frac{\hbar\,t}{m\,\sigma^2}\right)^2\left(1+{\cal B}\, \frac{m 
\,\lambda \,\sigma}{2\,\hbar^2}\right)}\,,
\ee
where ${\cal B} \approx 1.23$. For $\lambda=0$
one retrieves an expression similar to Eq. (\ref{GWP_variance}),
while in the limit  $\lambda \rightarrow \infty$, Eq. \eqref{variance_interactions}  diverges  for all  $t$ because 
the tail of the density matrix decays as $\propto \/x^2$,  even starting
from a Gaussian.

\begin{figure}[t]
\centering
\includegraphics{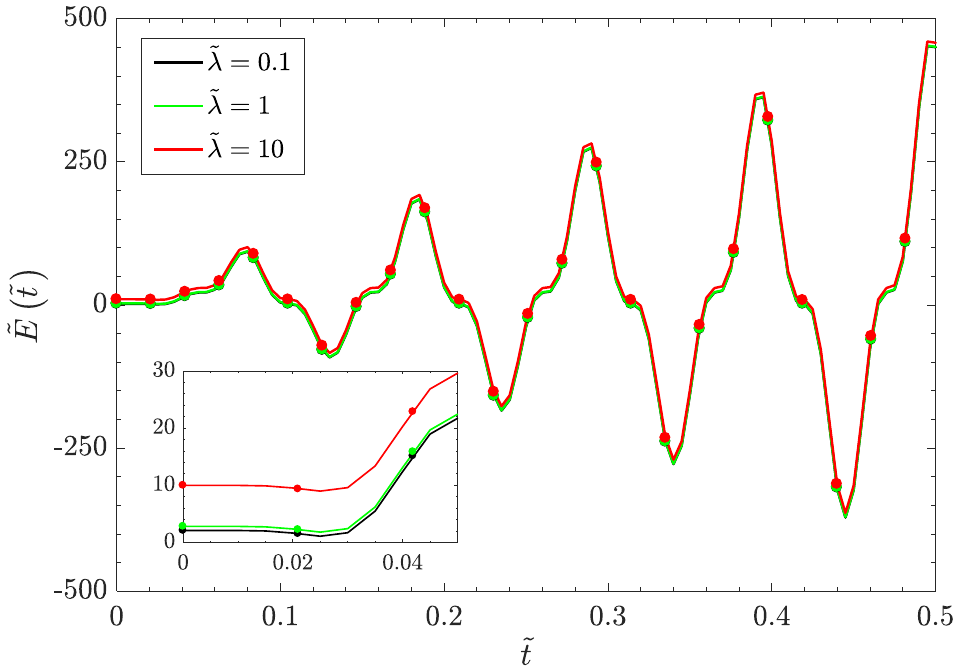}
\caption{Time evolution of the energy
  $\tilde{E} = \sqrt[3]{\frac{m}{\hbar^2 l}}\, E$ for
  two interacting particles subjected to a linear external potential:
  $x f(t)$, with driving function $f(t) = \ell \sin^3(\omega t)$.
  The system is prepared in the Gaussian wavepacket state
  (\ref{initial_WP_2b}). The  curves represent different
  values of  the parameter $\tilde{\lambda}$, which only shifts 
the total energy, as shown in the inset  for 
  short  times $\tilde{t}$ and different coupling strengths.
  The dots represent the energy values calculated from the numerical
  computation.} 
\label{energy_results_fig}
\end{figure}

As an additional 
check, we have calculated  numerically the total energy of a
two-particle system driven with $f(t) = \ell \sin^3(\omega t)$,
separating its center of mass and relative components. The analytical
value can be obtained from Eq. (\ref{energy_result_2b}),
and is represented by the solid lines in Fig. \ref{energy_results_fig}.
The dots represent  the values calculated numerically.
We have used $\tilde{\ell} = 200$, $\tilde{\omega}=60$,
$\tilde{\sigma}=2^{-1/2}$ and  $p_{0,j}=x_{0,j}=0$ for $j=1$, $2$. 
The interaction strengths,
$\tilde{\lambda} = 0.1$, $1$ and $10$, only displace  the
curves since their effects are encoded in the initial energy
factor $E(0)$ of Eq. (\ref{energy_result_2b}), as can be seen from
the inset of the plot. 
For this driving function we have
$f(nT)=\int_0^T f(\tau)\,d\tau =0$, therefore
from Eq. (\ref{energy_result_2b}) the energies at the stroboscopic times
are equal to the initial energy, \textit{i.e.} $E(nT) = E(0)$
for every $n$, and there is no heating of the system, in agreement with 
theoretical results  \cite{Sierra2015, He2019}
and experimental findings \cite{Pandey2019}.

\begin{figure}[t]
\centering
\includegraphics[width=151mm]{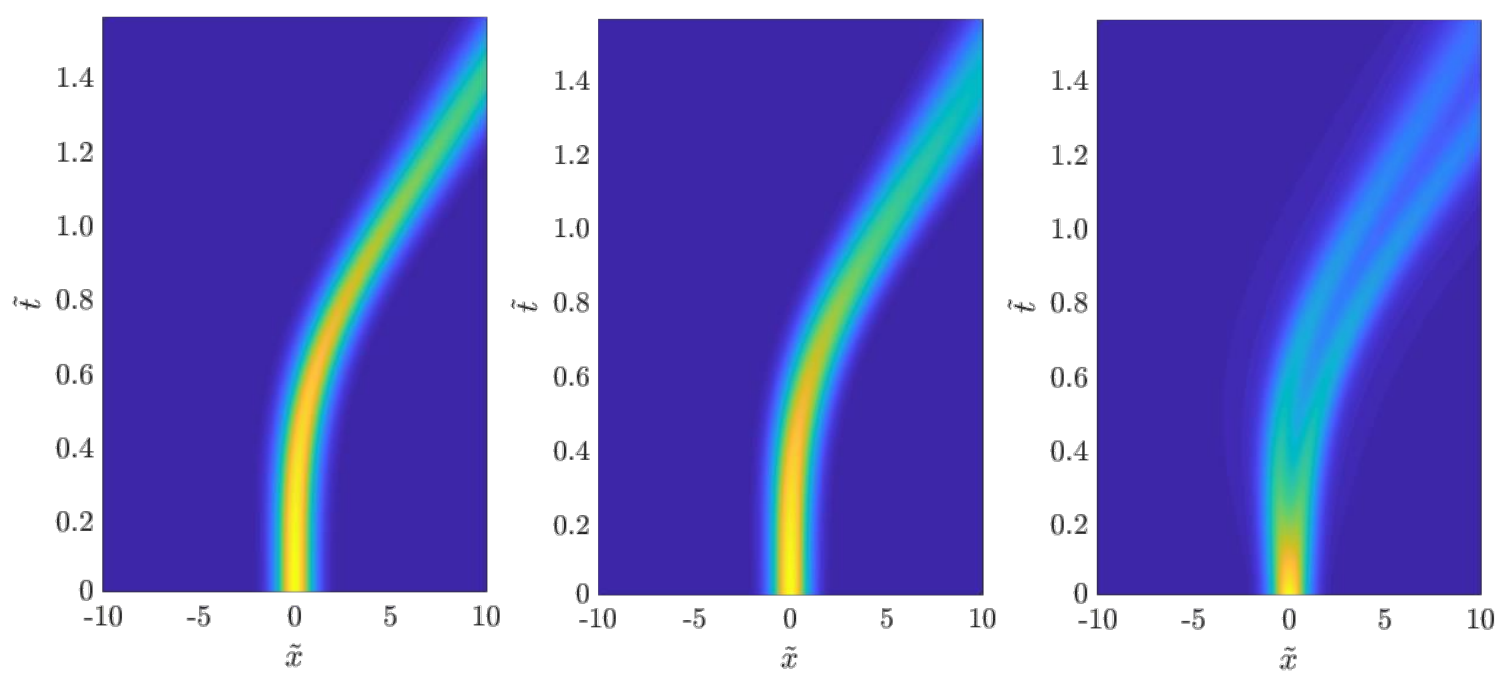}
\includegraphics[width=6.92mm]{Legend2b.pdf}
\caption{Evolution of density matrix (\ref{density_matrix_def}) for a Gaussian
  wavepacket (\ref{initial_WP_2b}) under the action of a linear external
  potential: $x f(t)$, where the driving function is
  $f(t) = \ell \left[\cos(\omega t)-1\right]$. Notice that the center
  of mass motion is uniformly accelerated to the right,
  as predicted analytically, and the wavepacket spreads over
  time as it would do for the undriven case. From left to right panels one has $\tilde{\lambda} = 0, 1, \infty$;
  moreover, $\tilde{\ell}=10$, $\tilde{\omega}=5$ and $\tilde{\sigma}=1$.}
\label{2b_ft_neq0}
\end{figure}

In the case where $\int_0^T f(\tau)\,d\tau\neq 0$,
we used Eq. (\ref{com_wavefunction_evol}) for the center of mass
initial wavefunction of Eq. (\ref{com_rel_wavepacket}), 
obtaining  the same result as when $\int_0^T f(\tau)\,d\tau =  0$, \textit{i.e.}
we retrieved Eq. (\ref{com_evolved}). For the relative motion
we have applied the same reasoning as before, by which we know
that the relative part of the wavepacket evolves according
to Eq. (\ref{propagator_eq}). We have performed a numerical simulation
of a system made of two $\delta$-interacting particles under the action
of a linear potential with driving function:
$f(t) = \ell \left[\cos(\omega t)-1\right]$. The results
for different interaction strengths $\lambda$ are reported in
Fig. \ref{2b_ft_neq0}, where the density matrices calculation
(\ref{density_matrix_calc}) is plotted, in correspondence of
$\tilde{\ell}=10$, $\tilde{\omega}=5$ and $\tilde{\sigma}=1$.
In this case the motion is uniformly accelerated to the right side
of the $x$-axis, indeed the translational parameter reads
$\xi(t) = \frac{\ell}{2m\omega^2} \left[\omega^2 t^2 -2 +
  2\cos(\omega t)\right]$. This has to be compared with the the case
$\int_0^T f(t) \,dt=0$, where the center of mass does not accelerate.

Concerning the spreading of the wavepacket, it is
the same as in the case without a driving potential and it also 
satisfies Eq. (\ref{variance_interactions}) with ${\cal B} \approx 1.23$.
In conclusion, there is no difference for the wavepacket spreading
between the results of a
driving function whose integral over a period vanishes  or not.

\section{Many-body problem}
The analysis done so far can be generalized to 
many-body systems with  $N$ interacting particles,  a generic interacting
potential $V_{2b}(x_j-x_i)$ and under the action of an external
linear time--dependent potential. The Schr\"odinger equation reads
\be
\label{schro_manybody_chi}
i\,\hbar\,\frac{\p \chi}{\p t}=\sum_{j=1}^N\left[-\frac{\hbar^2}{2\,m} \frac{\p^2}{\p x_j^2}+x_j \,f(t)\right]\chi+\sum_{j>i}V_{2b}(x_j-x_i) \chi\,.
\ee
Performing the translation and a gauge transformation 
\be
\label{multipart_states}
\chi(x_1,\dots,x_N,t)\equiv \prod_{j=1}^{N} e^{i \theta(x_j ,t)} \eta(y_1,\dots,y_N,t)\,,
\ee
the wavefunction $\eta(y_1,\dots,y_N,t)$ satisfies
the Schr\"odinger equation without the external driving, \textit{i.e.}
\be
\label{schro_manybody_eta}
i\,\hbar\,\frac{\p \eta}{\p t}=-\frac{\hbar^2}{2\,m} \sum_{j=1}^N \frac{\p^2 \eta}{\p y_j^2}+\sum_{j>i}V_{2b}(y_j-y_i) \eta\,,
\ee
where $y_j(t) = x_j - \xi(t)$, $\forall j$, 
therefore the interacting potential is invariant under
these transformations: $V_{2b}(y_j-y_i)=V_{2b}(x_j-x_i)$.

Using the initial conditions $\xi(0)=0$ and $\theta(x_j,0)=0$, $\forall j$,
the parameter $\xi(t)$ and the gauge phase
$\theta(x_j,t)$ satisfy Eqs. (\ref{transl_parameter}) and (\ref{theta_f(t)}).
Hence, the two wavefunctions coincide at initial time $t=0$.

The complete solution of the Schr\"odinger equation (\ref{schro_manybody_chi})
can be formally written as 
\be
\label{complete_sol_N}
\chi(x_1,\dots,x_N,t)\,=\,\prod_{j=1}^{N} \left[e^{i \theta(x_j ,t)} e^{-i\frac{\xi(t)}{\hbar}\hat{p}_j}\right] e^{-i\frac{t}{\hbar}\hat{H}_0}\chi(x_1,\dots,x_N,0)\,,
\ee
where the undriven Hamiltonian of one--dimensional many-particles
systems has  the general form
\be
\hat{H}_0 = \sum_{j=1}^N \frac{\hat{p}_j^2}{2m}+\sum_{j>i}V_{2b}(x_j-x_i)\,.
\label{gen_H0}
\ee
In (\ref{complete_sol_N}) the momentum operator $\hat{p}_j$
is the generator of the translation for the $j$-th particle, and $\eta$
is the solution of the Schr\"odinger equation with no linear driving.

The generalization of the two-body results for the expectation values
of physical observables is straightforward.
Firstly, we can compute the total energy of the system evaluating the
expectation value of the driven Hamiltonian. In the calculation we use the
conservation of the total momentum $\hat{P}=\sum_{j=1}^N \hat{p}_j$
for the undriven Hamiltonian $\hat{H}_0$, \textit{i.e.}
$\left[\hat{H}_0, \hat{P}\right]=0$, valid in the considered case
in which the interaction $V_{2b}$ depends on the relative distance between
the particles (see more comments in Section \ref{comments}).
Using the commutation relations   
we find for a general (also non-periodic) driving function $f(t)$:
\begin{eqnarray}
\label{energy_result_Nb}
E(t) &= & E(0)+\frac{N}{2m}\left[\int_0^t f(\tau)d\tau \right]^2+\sum_{j=1}^N p_{0,j} \left[\frac{t}{m}f(t) -\frac{1}{m}\int_0^t f(\tau)d\tau \right] + \\
&& - \frac{N f(t)}{m}\int_0^t d\tau \int_0^\tau f(\tau')d\tau' +\sum_{j=1}^N x_{0,j} \left[f(t)-f(0)\right]\,,\nonumber
\end{eqnarray}
which generalizes  Eq. (\ref{energy_result_2b}).
As for the two-body case, 
if $f(t)$ is periodic in time and its integral over  a
time--period vanishes, then the energy is conserved at
stroboscopic times if $f(t=0)=0$.
Once again, there is a decoupling between the interactions
and the external linear driving potential, since  the effect of the  
interactions among particles is encoded in the initial value of
the energy $E(0)$, while the remaining terms collect the effect of the 
external potential. 

Let us now focus on periodic driving functions. 
As before, we discuss separately the cases 
when  $\int_0^T f(\tau)\,d\tau=0$ and $\neq 0$.
In the first case, the gauge phase at stroboscopic times is
independent on the position variables, while the parameter $\xi$
is linear in the stroboscopic factor $n$, indicating a stroboscopic
motion with constant velocity. Using the fact that 
$\left[\hat{H}_0, \hat{P}\right]=0$
and the Baker-Campbell-Hausdorff formula on Eq. (\ref{complete_sol_N})
evaluated at  $t=nT$, we find the Floquet
Hamiltonian
\be
\label{floq_ham_N}
\hat{H}_F=\sum_{j=1}^N \left[\frac{\hat{p}_j^2}{2\,m}+\frac{\xi(T)}{T}\hat{p}_j-\hbar\frac{\theta(T) }{T}\right]+\sum_{j<i} V_{2b}(x_j-x_i)\,.
\ee
Hence, if the undriven Hamiltonian describes an integrable model,
also the Floquet Hamiltonian is exactly solvable since it has the
same two-body interaction potential among particles and presents
only a shift in the momenta. For the micro-motion operator one finds
\be
\label{micromotion_N}
\hat{U}_F(t,0) = e^{i t \sum_{j=1}^N \left[\frac{\theta\left(x_j,t\right)}{t}-\frac{\theta(T)}{T}\right]}\,e^{i\frac{t}{\hbar} \left[\frac{\xi(T)}{T}-\frac{\xi(t)}{t}\right]\sum_{j=1}^N \hat{p}_j}\,.
\ee
If $f(t)$ has a non-vanishing integral over a driving period, then the Floquet Hamiltonian reads 
\be
\hat{H}_F = \sum_{j=1}^N \left[\frac{\hat{p}_j^2}{2m}-\hbar\frac{\theta(x_j,T)}{T}\right]-\frac{N}{6m}\left[\int_0^T f(\tau)\,d\tau\right]^2 +\sum_{j<i} V_{2b}(x_j-x_i)\,,
\ee
which presents a time--independent $x$-linear potential term
acting on all the particles. In this case, as we saw for the one-body
problem, the system is governed by a stroboscopic dynamics with a uniform
acceleration, since the translational parameter depends quadratically on
the stroboscopic factor: $\xi(nT)\propto n^2$. The micro-motion operator reads:
\begin{eqnarray}
\nonumber
\hat{U}_F(t,0) &=& e^{\frac{i}{\hbar} \left\{ t\hbar\sum_{j=1}^N \left[\frac{\theta\left(x_j,t\right)}{t}-\frac{\theta\left(x_j,T\right)}{T}\right] -N \frac{t}{2m} \int_0^T f(\tau)\,d\tau \cdot \left[\frac{1}{3}\left(1+2\frac{t^2}{T^2}\right)\int_0^T f(\tau)\,d\tau \right]-N\xi(t)\frac{t}{T}\int_0^T f(\tau)\,d\tau \right\}}\cdot\\
 &&\cdot \,e^{-\frac{i}{\hbar} \left[\xi(t) + \frac{t^2}{2mT} \int_0^T f(\tau)\,d\tau\right]\sum_{j=1}^N \hat{p}_j}\,.
\end{eqnarray}

\subsection{Comments}
\label{comments}
We pause here to comment on the generality of our findings. The
main results in the case $\int_0^T f(\tau)\,d\tau=0$
are Eqs. (\ref{floq_ham_N}) and (\ref{micromotion_N}). They are valid
for any form of the two-body potential $V_{2b}$
and therefore for any interacting
Hamiltonian (\ref{gen_H0}), integrable or not. 
The crucial assumption we have made is that 
the two-body potential $V_{2b}$ depends only on 
the relative distance $x_i-x_j$, otherwise 
$V_{2b}(x_i,x_j)$ would be in general different from $V_{2b}(y_i,y_j)$
when the transformation $y_j=x_j-\xi(t)$ is done. Since $V_{2b}(x_j-x_i)=V_{2b}(y_j-y_i)$ then
the equations of motions for the wavefunction $\eta(y_1,\dots,y_N,t)$ are exactly the same
of those for the wavefunction $\chi(x_1,\dots,x_N,t)$, except for the fact that the time--periodic
linear potential has been removed. Notice, that in presence
of one-body potentials $V_{1b}(x_i)$, breaking translational invariance, this fact would be no
longer valid. When the interacting many-body Hamiltonian has only the kinetic
term plus a time--independent two-body potential $V_{2b}$ depending only on the relative distance
between the particles, then the conservation
of the total momentum of the undriven Hamiltonian $\hat{H}_0$ is guaranteed:
$$\left[\hat{H}_0, \hat{P}\right]=0\,,$$
a relation we subsequently used to determine the Floquet Hamiltonian, the micro-motion operator
and the expression of the energy at time $t$.  

We conclude that if, in addition, $\hat{H}_0$ turns out to be integrable, then the associated 
Floquet Hamiltonian is integrable too. We have presented the analysis for a many-body systems made of  bosons,
but it could  equally be 
applied to a many-body systems made of fermions or Bose--Fermi mixtures. In few words, our results are valid
for any one--dimensional integrable Hamiltonian in the \textit{continuum}. 
This also includes the Gaudin-Yang model
for one--dimensional Fermi gases, integrable Bose-Fermi mixtures,
integrable multi-component Lieb--Liniger Bose gases and Calogero-Sutherland
models (in the absence of external one-body harmonic potential)
\cite{Korepin1993,Sutherland04,Mussardo10,Gaudin14}.
     
Hence, having in mind the broad generality of our results, we shall present below a study of the paradigmatic 
Lieb--Liniger model driven by an external linear time--dependent potential whose driving function
has a vanishing integral over a driving period. 

\subsection{Driven Lieb--Liniger gas}
The Lieb--Liniger model describes a gas of $N$ bosons with
$\delta$-contact repulsive interactions in one--dimension
\cite{LiebLiniger1963}, tha is  $V_{2b}(x_j-x_i) = \lambda \delta(x_j-x_i)$,
with $\lambda>0$ the interaction parameter. The dynamics
of the Lieb--Liniger model in a linear potential was studied in
\cite{Jukic2010}, while we refer to \cite{Chen,Ablowitz2004} for a study
of the classical counterpart of the Lieb--Liniger model, the nonlinear
Schr\"odinger equation, in the presence of a time--dependent linear potential.  
The Floquet analysis
of the Lieb--Liniger model
with a periodic tilting was studied in \cite{ourPRL2019},
where it was discussed 
the stroboscopic evolution written in terms of the eigenfunctions of the
Floquet Hamiltonian in Eq. (\ref{floq_ham_N}). Here we make a further
step forward, giving a procedure for getting an expression for the
time evolution of a generic wavepacket. 

The undriven Hamiltonian of this system, \textit{i.e.}
\be
\hat{H}_0 = \sum_{j=1}^N \frac{\hat{p}_j^2}{2m} +
\lambda \sum_{j<i} \delta(x_j-x_i)
\,\,\,
\ee
is an integrable Hamiltonian and an
exact expression of its eigenfunction can be obtained  using the Bethe
ansatz technique \cite{Korepin1993,Yang1969}. Therefore we can write the
eigenfunctions for the Floquet Hamiltonian (\ref{floq_ham_N}) as Bethe
ansatz states
\be
\left| \tilde{u}\right\rangle = \sum_P A_P(Q)\, e^{\frac{i}{\hbar}\sum_{j=1}^N k_{P_j} x_j}\,,
\ee
where $Q$ is the permutation index which specifies the order of the
particles, while $P$ is the permutation index of the pseudo-rapidities $k_j$,
which are undetermined until boundary conditions are chosen
\cite{Korepin1993,Mussardo10} (we refer to \cite{ourPRL2019} for a
discussion on the relation between the boundary conditions and the external
linear potential). The amplitudes $A_P(Q)$ can be written as
\be
\nonumber
A_P = \mathcal{N}\,(-1)^P\,\prod_{j<l} \left(k_{P_j} -k_{P_l} +i\frac{m\lambda}{\hbar^2}\right)\,,
\ee
where $\mathcal{N}$ represents the normalization factor. The respective
quasi-energies are given by 
\be
\label{quasien_LL}
\mathcal{E}_F = \frac{\hbar^2}{2m}\sum_{j=1}^N k_j^2 +\hbar\frac{\xi(T)}{T} \sum_{j=1}^N k_j -N\hbar \frac{\theta(T)}{T}\,.
\ee
For convenience, we will indicate the state
$\tilde{u}$ as ${\rm BAS}(k_1,\dots,k_N)$, where
${\rm BAS}$ stands for \textit{Bethe Ansatz State}. In order to understand
what happens for the $N$--body case, it is convenient to start from the
two-body problem. In this case we can write \cite{FranchiniBook}
\be
\label{bas_2b}
{\rm BAS}(k_1,k_2) = g(x_1,x_2) \theta_H (x_2-x_1)+g(x_2,x_1) \theta_H (x_1-x_2)\,,
\ee
where $\theta_H(x)$ is the Heaviside step function, while
\be
\nonumber
g(x_1,x_2)=\left[i(k_1-k_2)-\frac{m\lambda}{\hbar^2}\right]\,e^{i\left(k_1 x_1+k_2 x_2\right)}+\left[i(k_1-k_2)+\frac{m\lambda}{\hbar^2}\right]\,e^{i\left(k_2 x_1+k_1 x_2\right)}\,.
\ee
Hence, $g(x_1+a,x_2+a) = g(x_1,x_2)\,e^{ia\left(k_1+k_2\right)}$ for generic $a$,
and the action of the micro-motion operator (\ref{micromotion_N}) on the
${\rm BAS}$  will give the following Floquet modes
\be
u(t) = {\rm BAS} \left(k_1-\frac{1}{\hbar}\int_0^t f(\tau)\,d\tau, k_2-\frac{1}{\hbar}\int_0^t f(\tau)\,d\tau)\right)\,e^{i\left(k_1+k_2\right)t\left[\frac{\xi(T)}{T}-\frac{\xi(t)}{t}\right]}\,e^{-i \left\{\frac{1}{m\hbar} \int_0^t d\tau\left[\int_0^\tau 
f(\tau')\,d\tau' \right]^2 +2\frac{t}{T} \theta(T)\right\}}\,.
\ee
Apart from a phase, the Floquet modes are then Bethe ansatz states
with shifted pseudomomenta. The Floquet states
from Eqs. (\ref{floq_states_def}) and (\ref{quasien_LL}), read
\be
 \psi_F(t)= {\rm BAS}\left(k_1-\frac{1}{\hbar}\int_0^t f(\tau)\,d\tau, k_2-\frac{1}{\hbar}\int_0^t f(\tau)\,d\tau)\right)\,e^{-i\left(k_1+k_2\right)\xi(t)}\,e^{-\frac{i}{m\hbar}\int_0^t d\tau \left[\int_0^\tau f(\tau')\,d\tau'\right]^2}\,,
\ee
and the total momentum expectation value of the Floquet states is therefore
$\left\langle \hat{P} \right\rangle_F(t) = \hbar\left(k_1+k_2\right)-\frac{2}{\hbar}\int_0^t f(\tau)\,d\tau$. These results may be easily extended
to the many-body case. The Floquet modes can be written as:
\begin{eqnarray}
 u(t) & = & {\rm BAS}\left(k_1-\frac{1}{\hbar}\int_0^t f(\tau)\,d\tau,\dots, k_N-\frac{1}{\hbar}\int_0^t f(\tau)\,d\tau)\right)\,e^{it\left[\frac{\xi(T)}{T}-\frac{\xi(t)}{t}\right]\sum_{j=1}^N k_j} \cdot \\
 && \cdot 
 \,e^{-i \left\{\frac{N}{2 m\hbar} \int_0^t d\tau\left[\int_0^\tau 
f(\tau')\,d\tau' \right]^2 +N\frac{t}{T} \theta(T)\right\}}\,,
\end{eqnarray}
while the Floquet states read
\be
 \psi_F(t)  = {\rm BAS}\left(k_1-\frac{1}{\hbar}\int_0^t f(\tau)\,d\tau, \dots,k_N-\frac{1}{\hbar}\int_0^t f(\tau)\,d\tau)\right)\,e^{-i\xi(t) \sum_{j=1}^N k_j}\,e^{-i\frac{N}{m\hbar}\int_0^t d\tau \left[\int_0^\tau f(\tau')\,d\tau'\right]^2}\,.
\ee
The total momentum of the Floquet states is then
\be
\left\langle \hat{P}\right\rangle_F (t) =\hbar \sum_{j=1}^N k_j - \frac{N}{\hbar} \int_0^t f(\tau)\,d\tau\,.
\ee
In particular one can calculate the time evolution of a
generic wavepacket for this system as 
\be
\chi(x_1,\dots,x_N,t)=\int A(k_1,\dots,k_N)\,\psi_F(t)\, d^N k\,,
\ee
which is an extension of the one-body equation (\ref{floq_generic_solution}).

It is worth stressing that this is a non-trivial expansion to evaluate:
Indeed, once the initial wavepacket has been chosen at $t=0$, one needs
to evaluate the time--independent amplitudes $A(k_1,\dots,k_N)$ inverting
the integral by multiplying by $\psi_F^*(t)$, and then
evaluate the $N$-dimensional integral on the right hand side. 

\section{Conclusions}
In this paper we have studied the effect of a time--dependent 
linear external potential on one--dimensional quantum systems
made of one-, two- and many-particles.
The potential could physically represent a
time varying gravitational linear force,
or a time varying electric field acting on the system,
therefore its analysis is interesting in many different contexts.
The key point of our approach has been to solve the problem for a
generic driving function by applying a gauge transformation
on the wavefunction and a translation over the position variables. Doing so,
we have been able to compute expectation values for different observables
such as the center of mass position of a wavepacket and its variance, and
the way these observables depend on time. We have observed that the
external driving does not affect the spread of a wavepacket,
which depends instead only on the interaction effects. This is the
result of the decoupling of the external potential
which takes place already from the two-particles case,
due to the linearity of the potential. This decoupling acts
at the level of the center of mass and relative coordinates and
can be observed also in the behaviour of the total energy of the system,
which oscillates in time depending on the form of the driving function $f(t)$.
We derived expressions for the energy of the state at any time also for
non-periodic driving function. The system in general does not conserve
the energy, apart from some specific cases, \textit{e.g.} if $f(t)$ is
constant in time. However, when $f$ is periodic in time
and its integral on a time--period
vanishes, plus $f(t=0)=0$, 
then the energy at stroboscopic times is conserved (notice that,
at stroboscopic times, the expectation value of the full
Hamiltonian does not need to be equal
to the expectation value of the Floquet Hamitonian).  
When $f$ is periodic, but its integral on a time--period
is non-vanishing, then the energy at stroboscopic times is in general
not conserved. 

For a periodic driving, we have analysed in
detail the dynamics of the systems. In this case
we have employed the Floquet approach and written down
the Floquet Hamiltonian and the micro-motion operator,
describing the time evolution of the system at stroboscopic times and
generic intermediate times, respectively. Our results, as discussed in Section \ref{comments}, are valid when
the two-body interaction terms depend only on the relative distance between
the particles so that the total momentum commutes with the undriven Hamiltonian.
If the undriven Hamiltonian is integrable, and obey such conditions,
then, when $\int_0^T f(\tau)\,d\tau=0$,
the Floquet Hamiltonian is integrable too.
Therefore, our results are valid
for any one--dimensional integrable Hamiltonian on the continuum
including the Gaudin-Yang model
for one--dimensional Fermi gases, integrable Bose-Fermi mixtures,
integrable multi-component Lieb--Liniger Bose gases and Calogero-Sutherland
models (in absence of external one-body harmonic potential).
It would be of interest to study
the integrablity of the Floquet Hamiltonian and the micro-motion operator
for undriven integrable lattice Hamiltonians subjected to time--periodic linear potentials
(or magnetic fields) suitably extending the method presented here.

If the integral of the driving function on
a period of oscillation is, on the contrary, non-vanishing,
then the Floquet Hamiltonian can be shown to be time--independent
and it contains a linear, constant in time, external potential.
In this case, such term can be eliminated using the same
recipe of a gauge transformation and a translation over the position
variables. The study whether
such Floquet Hamiltonians are in general formally integrable
is a very interesting
topic of future research.

We finally obtained expressions for the Floquet states for one-,
two- and many-body cases with contact interactions, where it has been
observed that they essentially retains the form of the eigenfunctions
of the original undriven Hamiltonian with a time dependent translation
over the momenta (or pseudo-momenta). Our approaches can be applied
to any many-body system where the particles interact with a two-body
potential which depends on the difference between particles positions and are
translationally invariant. It would be very
interesting to consider the effects of different boundary conditions
on the problem in finite-size systems, and employing a
Floquet engineering approach to study ac--Stark shifts and multiphoton
resonances \cite{Holthaus2016} for single and many-particles systems.

\vspace{0.5cm}

{\bf Acknowledgments} 

Discussions with G. Santoro, J. Schmiedmayer and M. Aidelsburger are gratefully acknowledged.
The authors thank the Erwin
Schr\"odinger Institute (ESI) in Wien for support during the Programme "Quantum Paths". 
GS acknowledges financial support from the Spanish grants PGC2018-095862-B-C21,
QUITEMAD+ S2013/ICE-2801, SEV-2016-0597 of the ''Centro de Excelencia Severo
Ochoa'' Programme and the CSIC Research Platform on
Quantum Technologies PTI-001.

\end{document}